\documentclass[twocolumn]{aastex631}

\usepackage{hyperref}
\usepackage[utf8]{inputenc}
\usepackage{amsmath}
\shorttitle{LOFAR explores the low-frequency radio emission of GASP jellyfish galaxies}
\shortauthors{Ignesti et al.}
\graphicspath{{./}{figures/}}

\begin{document}

\title{Walk on the Low Side: LOFAR explores the low-frequency radio emission of GASP jellyfish galaxies}

\correspondingauthor{Alessandro Ignesti}
\email{alessandro.ignesti@inaf.it}

\author[0000-0003-1581-0092]{Alessandro Ignesti}\affiliation{INAF-Padova Astronomical Observatory, Vicolo dell’Osservatorio 5, I-35122 Padova, Italy}

\author[0000-0003-0980-1499]{Benedetta Vulcani}\affiliation{INAF-Padova Astronomical Observatory, Vicolo dell’Osservatorio 5, I-35122 Padova, Italy}

\author[0000-0001-8751-8360]{Bianca M. Poggianti}\affiliation{INAF-Padova Astronomical Observatory, Vicolo dell’Osservatorio 5, I-35122 Padova, Italy}

\author[0000-0002-1688-482X]{Alessia Moretti}\affiliation{INAF-Padova Astronomical Observatory, Vicolo dell’Osservatorio 5, I-35122 Padova, Italy}

\author[0000-0001-5648-9069]{Timothy Shimwell}\affiliation{ASTRON, the Netherlands Institute for Radio Astronomy, Postbus 2, 7990 AA Dwingeloo, The Netherlands}\affiliation{Leiden Observatory, Leiden University, PO Box 9513, 2300 RA Leiden, The Netherlands}

\author[0000-0002-9325-1567]{Andrea Botteon}\affiliation{Leiden Observatory, Leiden University, PO Box 9513, 2300 RA Leiden, The Netherlands}

\author[0000-0002-0587-1660]{Reinout J. van Weeren}\affiliation{Leiden Observatory, Leiden University, PO Box 9513, 2300 RA Leiden, The Netherlands}
\author[0000-0002-0692-0911]{Ian D. Roberts}\affiliation{Leiden Observatory, Leiden University, PO Box 9513, 2300 RA Leiden, The Netherlands}

\author[0000-0002-7042-1965]{Jacopo Fritz}\affiliation{Instituto de Radioastronomía y Astrofísica, UNAM, Campus Morelia, A.P. 3-72, C.P. 58089, Mexico}

\author[0000-0002-8238-9210]{Neven Tomi\v{c}i\'{c}}\affiliation{INAF-Padova Astronomical Observatory, Vicolo dell’Osservatorio 5, I-35122 Padova, Italy}

\author[0000-0001-5766-7154]{Giorgia Peluso}\affiliation{INAF-Padova Astronomical Observatory, Vicolo dell’Osservatorio 5, I-35122 Padova, Italy}\affiliation{Dipartimento di Fisica e Astronomia ”G. Galilei”, Universit`a di Padova, Vicolo dell’Osservatorio 3, 35122 Padova, Italy}

\author[0000-0001-9143-6026]{Rosita Paladino}\affiliation{INAF, Istituto di Radioastronomia di Bologna, via Piero Gobetti 101, 40129 Bologna, Italy}

\author[0000-0002-0843-3009]{Myriam Gitti}\affiliation{Dipartimento di Fisica e Astronomia, Università di Bologna, via Piero Gobetti 93/2, 40129 Bologna, Italy}\affiliation{INAF, Istituto di Radioastronomia di Bologna, via Piero Gobetti 101, 40129 Bologna, Italy}

\author[0000-0001-9184-7845]{Ancla M\"uller}\affiliation{Ruhr University Bochum, Faculty of Physics and Astronomy, Astronomical Institute, Universit\"atsst 150, 44801 Bochum, Germany}

\author[0000-0003-3255-3139]{Sean McGee}\affiliation{School of Physics and Astronomy, University of Birmingham, Birmingham B15 2TT, United Kingdom}

\author[0000-0002-7296-9780]{Marco Gullieuszik}\affiliation{INAF-Padova Astronomical Observatory, Vicolo dell’Osservatorio 5, I-35122 Padova, Italy}

\begin{abstract}
Jellyfish galaxies, characterized by long filaments of stripped interstellar medium extending from their disks, are the prime laboratories to study the outcomes of ram pressure stripping. At radio wavelengths, they often show unilateral emission extending beyond the stellar disk, and an excess of radio luminosity with respect to that expected from their current star formation rate. We present new 144 MHz images provided by the LOFAR Two-metre Sky Survey for a sample of six galaxies from the GASP survey. These galaxies are characterized by a high global luminosity at 144 MHz ($6-27\times10^{22}$ W Hz$^{-1}$), in excess compared to their ongoing star formation rate. The comparison of radio and H$\alpha$ images smoothed with a Gaussian beam corresponding to $\sim$10 kpc reveals a sub-linear spatial correlation between the two emissions with an average slope $k=0.50$. In their stellar disk we measure $k=0.77$, which is close to the radio-to-star formation linear relation. We speculate that, as a consequence of the ram pressure, in these jellyfish galaxies the cosmic rays transport is more efficient than in normal galaxies. Radio tails typically have higher radio-to-H$\alpha$ ratios than the disks, thus we suggest that the radio emission is boosted by the electrons stripped from the disks. In all galaxies, the star formation rate has decreased by a factor $\leq10$ within the last $\sim10^8$ yr. The observed radio emission is consistent with the past star formation, so we propose that this recent decline may be the cause of their radio luminosity-to-star formation rate excess.
\end{abstract}


\section{Introduction} \label{sec:intro}

Star formation plays a key role in galaxy evolution, and understanding the physical effects that impact the properties and the distribution of star formation and/or inter-stellar medium (ISM) are one of the main focusses of modern, extra-galactic research. Amongst the processes able to alter the gas content of a galaxy, and hence quench its star formation, ram pressure stripping (RPS) is considered to be the most efficient mechanism in removing the ISM from galaxies in galaxy clusters \citep{Boselli2006,Boselli_2021}. The most extreme examples of galaxies undergoing strong RPS are the so-called jellyfish galaxies \citep{Fumagalli2014, Smith2010, Ebeling2014,Poggianti2017}. In the optical/UV band, these objects show extraplanar, unilateral debris extending beyond their stellar disks, and striking tails of ionised gas. Jellyfish galaxies mostly reside in galaxy clusters and are a transitional phase between infalling star-forming spirals and quenched cluster galaxies, hence they provide a unique opportunity to understand the impact of gas removal processes on galaxy evolution. Moreover,  in jellyfish galaxies we can study the physics of the interplay between stripped, cold ($<10^4$ K) ISM and the surrounding, hot ($\sim10^7$ K) intracluster medium (ICM), hence obtain insights into the more general astrophysical case of the interplay between cold clouds and hot winds. \\

A number of jellyfish galaxies are also found to show extraplanar, nonthermal radio emission \citep[e.g.,][]{Gavazzi_1978,Gavazzi_1986,Gavazzi1995,Gavazzi_1999, Vollmer_2013,Poggianti_2019,Ignesti_2020b, Chen_2020,Muller_2021,Roberts_2021a,Roberts_2021b,Ignesti_2021,Roberts_2021c,Luber_2022}. In general, radio continuum emission in spiral galaxies is composed of the thermal emission from the $\sim10^4$ K plasma in the HII regions produced via thermal bremsstrahlung, and the non-thermal synchrotron emission of the relativistic cosmic rays electrons (CRe), which are accelerated by supernovae shocks reaching energies of a few GeV \citep[e.g.][for a review]{Condon_1992}. Given the connection between supernovae and CRe and the fact that galaxies are generally optically-thin to radio wavelengths, the continuum radio emission is a reliable proxy to evaluate their star formation rate (SFR) \citep[e.g.,][]{Kennicutt_2012}.\\

At GeV energies, CRe energy losses are dominated by synchrotron emission in the galactic magnetic fields with typical values of 10-30 $\mu$G \citep[e.g.,][and references therein]{Beck_2005}, which results in nonthermal radio emission in the 100 MHz-10 GHz band, and Inverse Compton scattering with the galactic and stellar radiation field  \citep[e.g.,][]{Pacholczyk_1970,Longair_2011}. The CRe synchrotron lifetime, $t_{\text{syn}}$, determined by the synchrotron and, in a lesser extent, by the Inverse Compton losses can be expressed as:
\begin{equation}
\frac{t_{\text{syn}}}{\text{Myr}}\simeq34.2\left(\frac{\nu}{\text{GHz}} \right)^{-0.5}\left(\frac{B}{10\text{ }\mu\text{G}} \right)^{-1.5}\left(1+\frac{U_{\text{Rad}}}{U_{\text{B}}} \right)^{-1}\text{,}
\label{t-syn}    
\end{equation}
where $\nu$ is the emission frequency, $B$ is the magnetic field intensity, and $U_{\text{Rad}}/U_{\text{B}}$ is the ratio between the radiation field and magnetic field energy densities \citep[][]{Heesen_2016}. In the densest region of the galaxies additional losses can also occur  via bremsstrahlung and, for CRe with energy $\lesssim$1 GeV (i.e., those responsible for the radio emission below $\sim200$ MHz in a 10 $\mu$G magnetic field), via ionization captures \citep[e.g.,][]{Murphy_2009b, Basu_2015}. Therefore, the CRe radiative time depends on the combination of the aforementioned physical effects, and  on  the CR energy, the magnetic and radiation fields intensity, and the local ISM density \citep[for a detailed discussion, see][]{Murphy_2009b, Lacki_2010, Basu_2015}. So CRe emitting at different frequencies have different radiative time, with the CRe emitting in the 100-1000 MHz band being in the order of $10^{7-8}$ yr \citep[e.g.,][]{Basu_2015,Heesen_2019}. Within this timescale, CRe can move within the galaxies resulting in extended radio emission. The CRe transport can take place either via diffusion along and across magnetic field lines, streaming, or advection. The dominant mechanism has been observed to vary in the different galactic environments \citep[e.g.,][for a recent review]{Heesen_2021}. In the stellar disk, the transport is generally dominated by diffusion. Outside of the stellar disk, such as in the galactic radio halos, it is generally dominated by advection/streaming where, in the case of normal spiral galaxies, the CRe speed should be of the order of the Alfv\'en speed ($V=B/\sqrt{4\pi\rho}$, with $B$  magnetic field intensity, and $\rho$ medium mass density). In the case of jellyfish galaxies' tails, the CRe transport mechanism is advection by the ram pressure wind \citep[][]{Murphy2009,Vollmer_2013}. The CRe velocity should, therefore, be similar to the velocity of the stripped clouds which, depending on the galactic velocity and the distance from the disk, can be of the order of several hundreds of kilometers per second  \citep[][]{ Tonnesen_2021}. Hence the radio tail length, in absence of CRe re-acceleration/injection within the tail, should depend on the advection speed and the cooling time-scale \citep[e.g.,][]{Chen_2020, Muller_2021}, resulting in a dependence of the radio tail's length on the observed frequency and, specifically, in longer radio tails at lower frequencies \citep[][]{Ignesti_2021}. \\

High resolution observations have revealed radio-deficit regions on the leading edge of jellyfish galaxies  \citep[][]{Murphy2009}. Nevertheless, these systems, as well as spiral galaxies in clusters, have been reported to show a global excess of radio emission with respect to their current star formation rate when compared to similar objects outside of clusters  \citep[e.g.,][]{Gavazzi_1986,Niklas_1995,Gavazzi_1999,Boselli2006,Murphy2009,Vollmer_2010,Vollmer_2013,Chen_2020,Roberts_2021a}. This evidence has been generally explained as either being due to the compression, and resulting amplification, of the magnetic field frozen within the ISM during RPS \citep[e.g.,][]{Boselli2006,Boselli_2021,Farber_2022}, or as a consequence of the interactions between the CRe and the ICM-driven `shocklets'. Recently, \citet[][]{Ignesti_2021} suggested that, instead, it could be due to a fast decline of the SFR within the CRe radiative time which can leave a population of old CRe which dominates the global, nonthermal radio emission, and naturally causes a discrepancy between the radio luminosity and the present SFR.\\

Observing nonthermal radio emission extending beyond their disk proves that jellyfish galaxies host large-scale, extraplanar magnetic fields. Previous works constrained such magnetic fields around $\sim10-20$ $\mu$G in the stellar disks and $\sim4-6$ $\mu$G in the tails of two jellyfish galaxies \citep[JO206 and JW100, see][]{Muller_2021,Ignesti_2021}. Such fields can affect the geometry of tails \citep[e.g.,][]{Tonnesen_2014,Ruszkowski_2014,Ramon-Martinez_2018}, as well as the evolution of the stripped gas. Indeed, simulations of cold gas clouds subjected to a hot wind found that magnetic fields can have a significant impact on the dynamics of wind-cloud interactions by suppressing fluid instabilities and, thus, extending the lifetime of the gas and the star-forming clouds in the stripped gas \citep[e.g.,][]{Berlok2019,Cottle_2020, Sparre_2020}. This physical scenario is similar to jellyfish galaxies moving in the hot ($kT\simeq 3-8$ keV) ICM, thus it suggests that magnetic fields could be crucial to preserve the cold ($kT< 0.1$ keV), stripped ISM clouds in the hot, hostile environment by damping the thermal conduction between the two gasses. These cold clouds could further cool down to temperatures that allow new star formation episodes tracked by both the CO and H$\alpha$ emission \citep[e.g.,][]{Poggianti_2019,Moretti_2020}. The effect of the magnetic fields can potentially last for hundreds of Myr, thus permitting to the stripped ISM to survive in form of cold clouds in the ICM \citep[e.g.,][]{Ge_2021}. \\

 The scale length of jellyfish galaxies' magnetic fields is also thought to be increased by their interaction with the ICM. A galaxy moving in a magnetised medium, such as the ICM \citep[e.g.][]{Carilli_2002}, can sweep up the surrounding magnetic field. This process, which is known as `magnetic draping' \citep[e.g.][]{Dursi_2008,Pfrommer_2010} can produce a magnetic sheath that surrounds the ISM and extends into the galaxy's wake along the tail of the jellyfish galaxy. In this scenario, as a consequence of the injection of relativistic electrons by supernovae into an ordered magnetic field, jellyfish tails should show, in the absence of depolarization effects,  highly-polarized synchrotron emission \citep[][]{Sparre_2020}. Indeed, deep Very Large Array observations recently provided the first detection of these aligned magnetic fields in the tail of the jellyfish galaxy JO206 \citep[][]{Muller_2021}. \\

\begin{table*}[t]
    \centering
    \begin{tabular}{c|cccc|cc|ccc}
    \toprule
         Galaxy&$z$& log $M_{*}$ &Cluster&$z_{\text{cluster}}$&Beam &RMS &$S_{\text{Rad}}$ &$L_{\text{Rad}}$&SFR\\
         &&[$M_\odot$]&&&[arcsec$^2$]&[$\mu$Jy beam$^{-1}$]&[mJy]&[10$^{22}$ W Hz$^{-1}$]&[$M_\odot$ yr$^{-1}$]\\
         \hline
         JW39&0.0663&11.21&A1668&0.0634&10.7$\times$6.6&183&19.2$\pm$3.2&22.8$\pm$3.8&3.6$\pm$0.7\\
         J085&0.0355&10.67&A2589&0.0422&11.5$\times$5.2&109&56.4$\pm$9.6&27.0$\pm$4.6&6.0$\pm$1.0\\
         JO60&0.0622&10.40&A1991&0.0586&10.8$\times$5.5&145&14.7$\pm$2.2&12.0$\pm$1.8&4.5$\pm$0.9\\
         JW100&0.0619&11.47&A2626&0.0551&8.8$\times$6.4&94&13.8$\pm$2.0&10.2$\pm$1.5&4.0$\pm$0.8\\
         JO49&0.0451&10.68&A168&0.0452&13.3$\times$6.0&299&10.7$\pm$2.3&5.6$\pm$1.2&1.4$\pm$0.3\\
         JO206&0.0511&9.52&IIZW108&0.0489&11.9$\times$5.4&298&18.8$\pm$3.7&10.6$\pm$2.3&5.0$\pm$1.0\\
       
    \hline
    \end{tabular}
     \caption{\label{sample}Properties of the GASP-LOFAR sample. From left to right: GASP name; galaxy redshift; stellar mass from \citet[][]{Vulcani2018}; Host cluster name and  redshift \citep[][]{Gullieuszik_2020}; resolution and RMS of the 144 MHz images reported in Figure \ref{radio-halpha}; integrated radio flux density and luminosity at 144 MHz; total SFR from \citet[][]{Vulcani2018}. The integrated radio flux densities were measured within the last contour reported in Figure \ref{radio-halpha}. We report, from top to bottom, an alternative name for each galaxy: IC 4141, UGC 12582, 2MASX J14535156+1839064, IC 5337, 2MASX J01144386+0017100, WINGS J211347.41+022834.9 (source \url{http://cdsportal.u-strasbg.fr/}). }
\end{table*}
In this work we explore the nonthermal side of jellyfish galaxies by analyzing the low-frequency radio emission of a sub-sample of objects from the GASP survey \citep[GAs Stripping Phenomena in galaxies with MUSE\footnote{\url{https://web.oapd.inaf.it/gasp/}},][]{Poggianti2017}. The GASP sample consists of 114 galaxies in different environments and characterized by a range of stellar masses from $10^9$ to $3.2\times10^{11}$ M$_\odot$ with redshift $0.04<z<0.07$. The GASP project aims to investigate the process of gas removal in cluster galaxies by means of multi-wavelength studies. In this context, low-frequency ($<200$ MHz) observations represent an unique probe. On one hand, at these frequencies the radio emission is usually dominated by the nonthermal component \citep[e.g.,][]{Condon_1992,Tabatabaei_2016}. Hence low-frequency observations are a solid proxy of the star formation \citep[e.g.,][]{Gurkan_2018, Heesen_2019}. On the other hand, low-frequency ($\sim100$ MHz) observations probe low-energy particles that, due to their lower emissivity, have typical life-times of a factor $\sim3$ longer than the high-energy particles traced by high-frequency ($>$1 GHz) radio emission \citep[e.g.,][]{Pacholczyk_1970}. This entails that low-frequency observations probe the oldest electrons moving along the extraplanar magnetic fields and provide us the most comprehensive view of this phenomenon. For these reasons, we studied a sample of GASP galaxies with the LOw Frequency ARray \citep[LOFAR,][]{vanHaarlem_2013}, that, up to date, provides the highest sensitivity and resolution at these frequencies. Previous studies of jellyfish galaxies with LOFAR are presented in \citet[][]{Ignesti_2020b,Roberts_2021a,Roberts_2021b,Roberts_2021c,Ignesti_2021}. In the latter, a multi-frequency analysis of the GASP galaxy JW100 unveiled a connection between the star formation history and the nonthermal synchrotron emission.\\

We pursue these studies by extending them on a sample of 6 GASP galaxies in the northern sky. We present the new LOFAR data and a series of diagnostic diagrams that we designed to characterize the local connection between the radio emission, the SFR, and the properties of the stripped gas. In this paper we use a \cite{Chabrier2003} initial mass function (IMF) and the standard concordance 
cosmology parameters $H_0 = 70 \, \rm km \, s^{-1} \, Mpc^{-1}$, ${\Omega}_M=0.3$
and ${\Omega}_{\Lambda}=0.7$. At the galaxies redshift it results in $1''\simeq0.9-1.1$ kpc. We define the synchrotron spectrum as $S\propto\nu^{\alpha}$, where $S$ is the flux density at the frequency $\nu$, and $\alpha$ is the spectral index.
\section{Data analysis}

\subsection{The GASP-LOFAR sample}
\label{gl-sample}
We considered a sub-sample of 6 cluster galaxies from the GASP survey (Table \ref{sample}). This selection was necessarily driven by the declination of the targets. Indeed, LOFAR can effectively observe at high sensitivity only the northern sky where only 13 out of 114 GASP galaxies are located (Poggianti et al., in preparation). 
In this work we present 6 of those galaxies observed which have been observed by LoTSS. For two galaxies of our sample, JW100 and JO49, the LOFAR images have been already presented in  \citet[][]{Ignesti_2020b,Ignesti_2021} and \citet[][]{Roberts_2021a} respectively.\\

The galaxies of the GASP-LOFAR sample are characterized by extraplanar H$\alpha$ emission produced by warm filaments of stripped ISM \citep[][]{Poggianti2019}. We observe galaxies where the H$\alpha$ emission, in projection, is mostly contained within the disk, as well as examples of truncated disks, long tails, or unwinding arms. The sample includes both galaxies at an early and an advanced stage of stripping. According to the SFR-stellar mass relation \citep[][]{Vulcani_2018c}, half of the sample (JW100, JW39 and JO49) shows a deficiency in SFR that indicates an advanced stage of stripping. Finally, the sample also presents a high incidence of AGN with 5 out of 6 galaxies showing nuclear activity according to the Baldwin, Phillips \& Terlevich diagnostic diagrams \citep[BPT,][]{Baldwin1981}: JW39, JO85, JO49 \citep[][]{Peluso_2021} host LINER-like nuclei, while JW100 and JO206 have a Seyfert nucleus \citep[][]{Poggianti2017b, Radovich2019}.  
\subsection{Data preparation}
\subsubsection{LOFAR}
\label{lofi}
The 120-168 MHz images presented in this work are provided by the LOFAR Two-metre Sky Survey \citep[LoTSS, ][]{Shimwell_2017,Shimwell_2019,Shimwell_2022}. For details on the observation strategy and calibration of the LoTSS data we refer to \citet[][]{Shimwell_2022}. Following the LoTSS procedure, the datasets were calibrated by using the direction-dependent calibration and imaging pipeline {\scshape ddf-pipeline}\footnote{\url{ https://github.com/mhardcastle/ddf-pipeline}} v. 2.2. This pipeline was developed by the LOFAR Surveys Key Science Project and it corrects for ionospheric and beam model errors in the data. The latest version of the pipeline is described in \citet[][]{Tasse_2021}. The entire data processing procedure makes use of {\scshape prefactor}  \citep[][]{vanWeeren_2016,Williams_2016,deGasperin_2019}, {\scshape killMS} \citep[][]{Tasse_2014,Smirnov_2015} and {\scshape DDFacet} \citep[][]{Tasse_2018}. The observations presented here have been further processed to improve the calibration in a smaller region of the LOFAR field-of-view containing the target of interest, employing the "extraction and self-calibration" procedure described in \citet[][]{vanweeren_2021}. We then produced the final images of the galaxies at a central frequency of the observation, 144 MHz, by using {\scshape WSClean}  v2.10.1  \citep[][]{Offringa_2014} and using a ROBUST=-0.5 \citep[][]{Briggs_1995}. The imaging was carried out on the HOTCAT cluster \citep[][]{Taffoni_2020,Bertocco_2020}. The Root Mean Square (RMS) map noise of each map was measured in source-free regions of the sky and it is reported in Table \ref{sample}.  \\
 
In order to explore the low-brightness, extraplanar emission, in this work we considered the radio surface brightness above the $2\times$RMS level similarly to \citet[][]{Roberts_2021a}. Our choice is justified by the absence of large-scale artifacts at a similar level of brightness near the galaxies. This assured us that the $2\times$RMS level structures observed extending from the galaxies in the direction of the H$\alpha$ tails are not calibration or imaging artifacts. For JW100, due to the proximity with the bright Kite radio source \citep[see][]{Ignesti_2020b,Ignesti_2021}, to not include possible artifacts we preferred to conservatively cut the radio emission at the $3\times$RMS level. In the following analysis of the radio emission we adopted a calibration error of $20\%$, according to what is done in LoTSS \citep[][]{Shimwell_2019,Shimwell_2022}. In our calculation we neglect the contribution of the thermal radio emission because, at these frequencies, it corresponds to the $2-5\%$ of the total radio emission \citep[e.g.,][]{Tabatabaei_2017,Ignesti_2021} and it cannot be differentiated from the nonthermal emission with these data alone. 
 \subsubsection{MUSE}
 \label{muse}
We make use of the previous results obtained by the GASP team from the Multi Unit Spectroscopic Explorer (MUSE) optical observations of the sample. The MUSE observations, data reduction and the methods of analysis are described in \citet{Poggianti2017}. Briefly, data were reduced with the most recent available version of the MUSE pipeline and data cubes were averaged filtered in the spatial direction with a 5 × 5 pixel kernel ($\sim$1 arcsec$^2$).
Data cubes were corrected for extinction due to the dust in the Milky Way, using the value estimated at the galaxy position \citep[][]{Schlafly2011} and assuming the extinction law from \citet[][]{Cardelli1989}. Emission-only data cubes were obtained by subtracting the stellar-only component of each spectrum derived with our spectrophotometric code SINOPSIS \citep[SImulatiNg OPtical Spectra wIth Stellar population models,][\footnote{\url{https://www.irya.unam.mx/gente/j.fritz/JFhp/SINOPSIS.html}}]{Fritz2017}.
This code searches the combination of Single Stellar Population (SSP) model spectra that best fits the equivalent widths of the main lines in absorption and emission and the continuum at various wavelengths and provides spatially resolved estimates of a number of stellar population properties, such as stellar masses and star formation histories. In particular, it
provides information on the  average SFR in twelve age bins. These bins can be combined into larger bins in such a way that the differences between the spectral characteristics of the stellar populations are maximal \citep{Fritz2017}. \\

Emission-line fluxes and errors were derived using the IDL software KUBEVIZ \citep[][]{Fossati2016}. Only spaxels with signal-to-noise ratio S/N(H$\alpha$) $>$ 3 were  considered as reliable. H$\alpha$ luminosities were corrected both for stellar absorption and for dust extinction. The latter was estimated from the Balmer decrement assuming an intrinsic  value H$\alpha$/H$\beta$=2.86 and the \citet[][]{Cardelli1989} extinction law. SFRs were computed from the dust- and absorption-corrected H$\alpha$ luminosities  \citep[][]{Vulcani2018}, following the \citet[][]{Kennicutt1998a} relation:

\begin{equation}
    \text{SFR}=4.6\times10^{-42}\left(\frac{L_{H\alpha}}{\text{erg s$^{-1}$}} \right) \text{ $M_\odot$ yr$^{-1}$}.
    \label{kenny}
\end{equation}
In order to characterize the ionization mechanism of warm gas emitting in H$\alpha$, we consider both the  BPT diagnostic diagrams presented in \citet[][]{Radovich2019, Poggianti2019,Peluso_2021} (using the NII lines for every galaxy with the only exception of JW100, for which we use the SII lines because the nitrogen line was contaminated by a sky line).\\

\section{Results}
\subsection{The new LOFAR images}
\label{new_res}
\begin{figure*}[t]
\includegraphics[width=.475\linewidth]{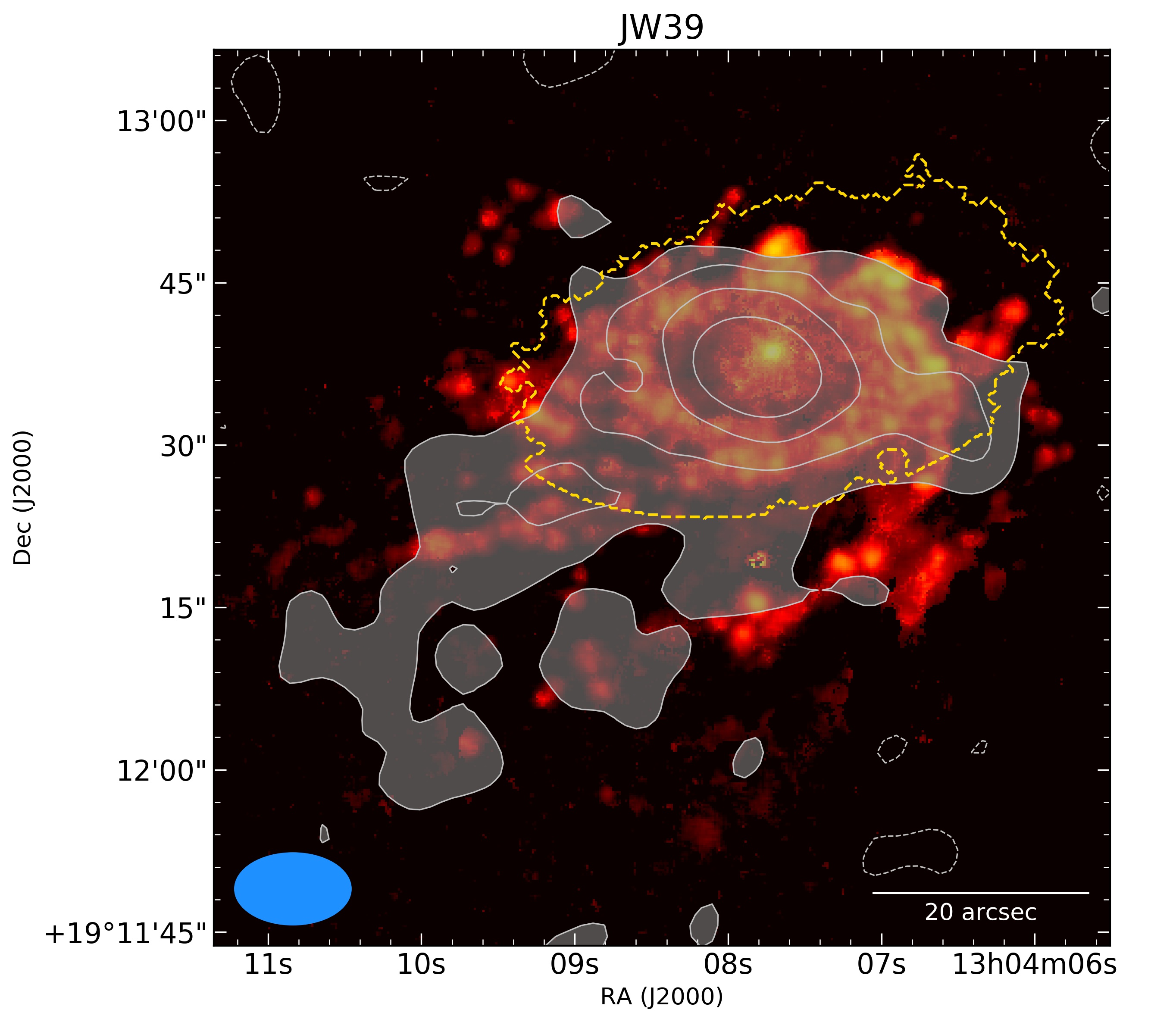}
\includegraphics[width=.475\linewidth]{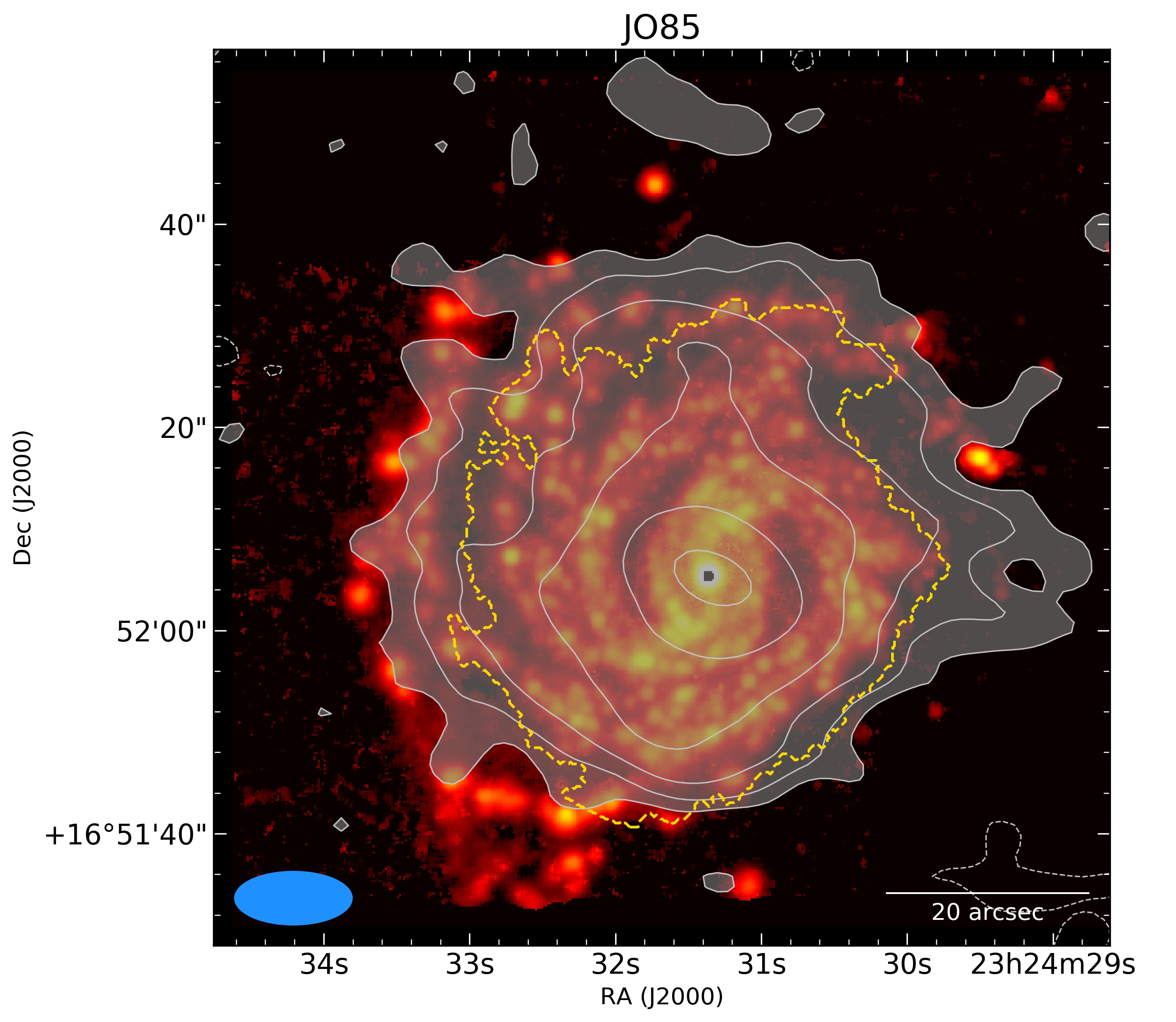}
\includegraphics[width=.475\linewidth]{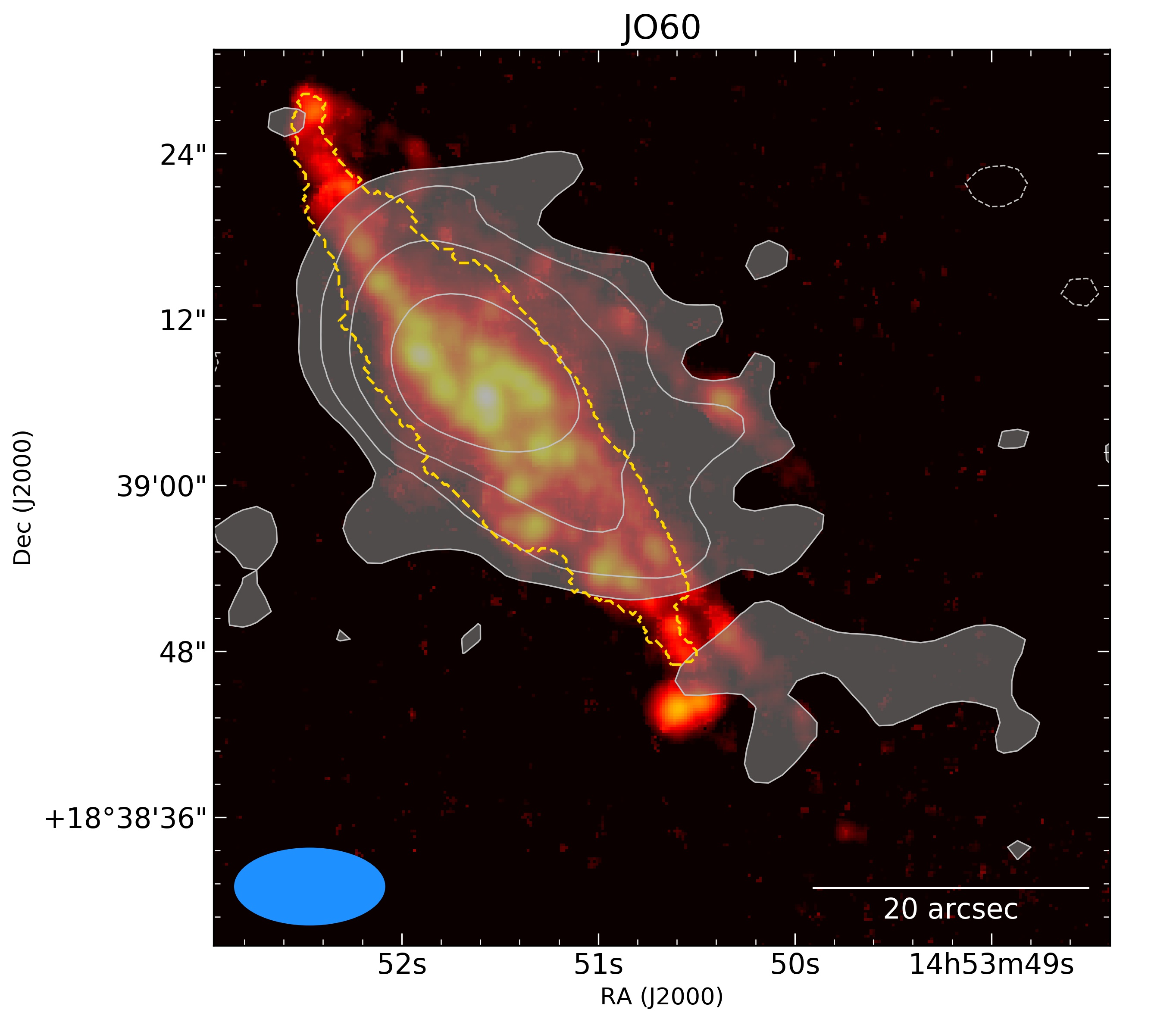}
\includegraphics[width=.475\linewidth]{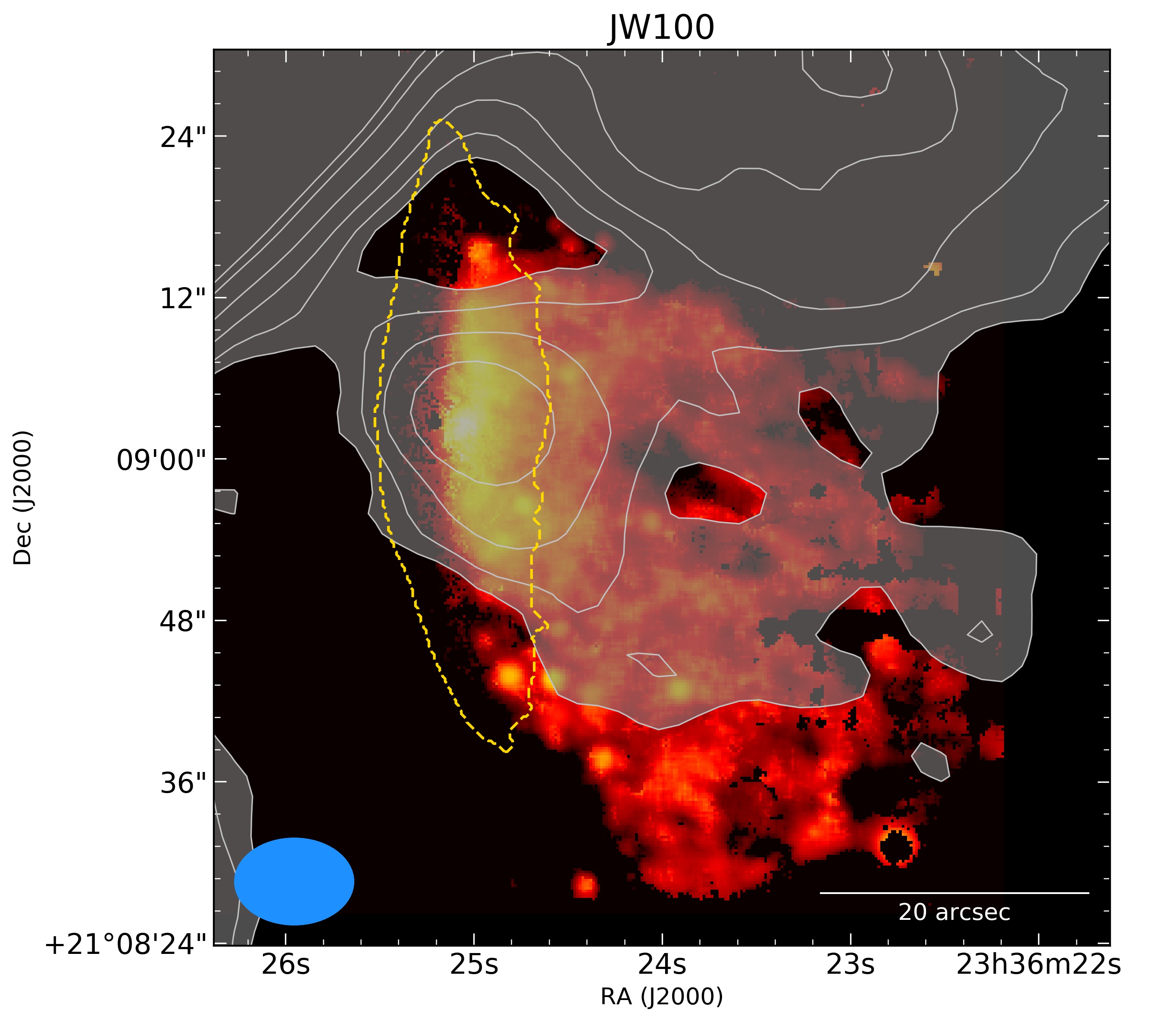}
\includegraphics[width=.475\linewidth]{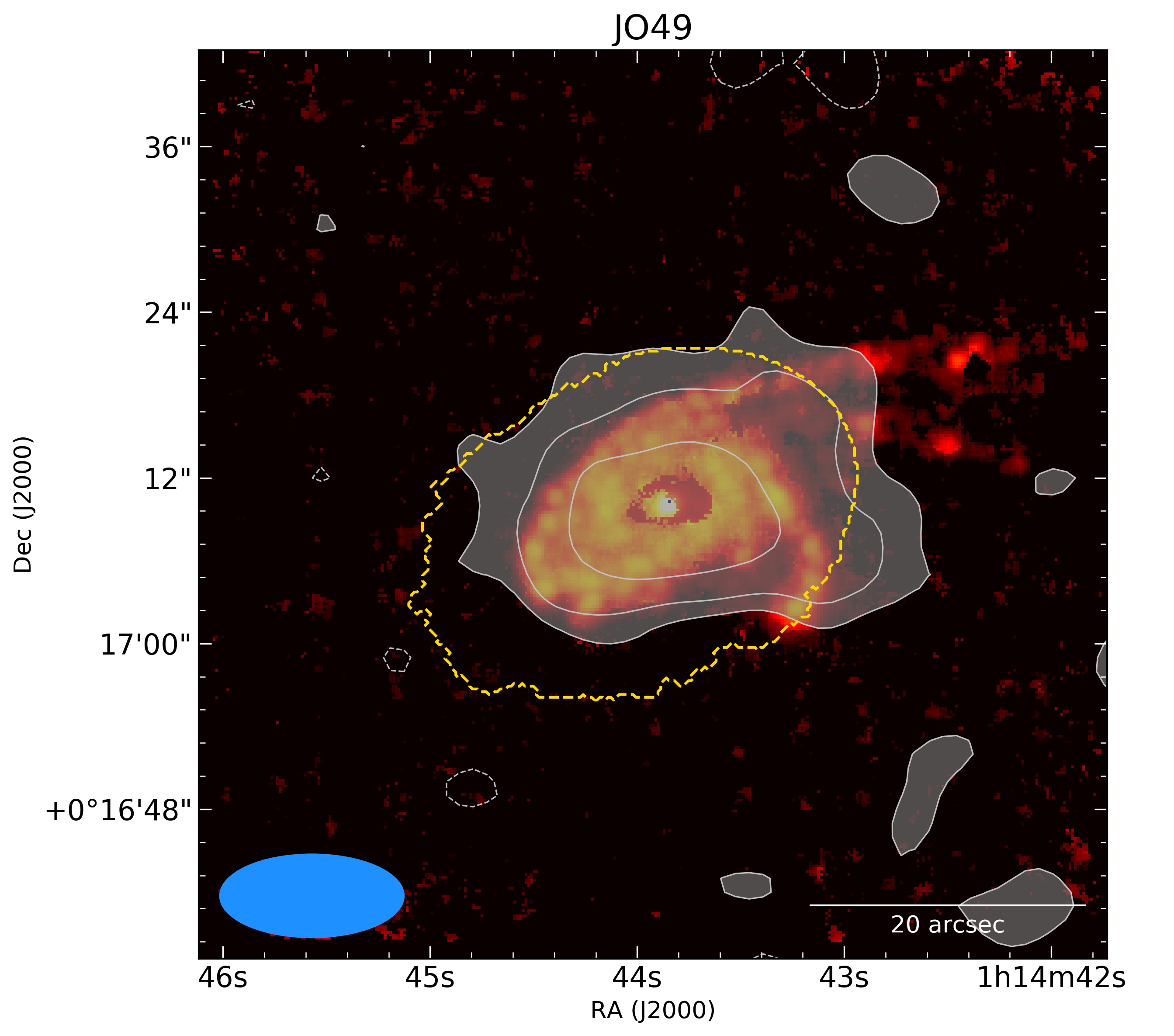}
\hfill\includegraphics[width=.475\linewidth]{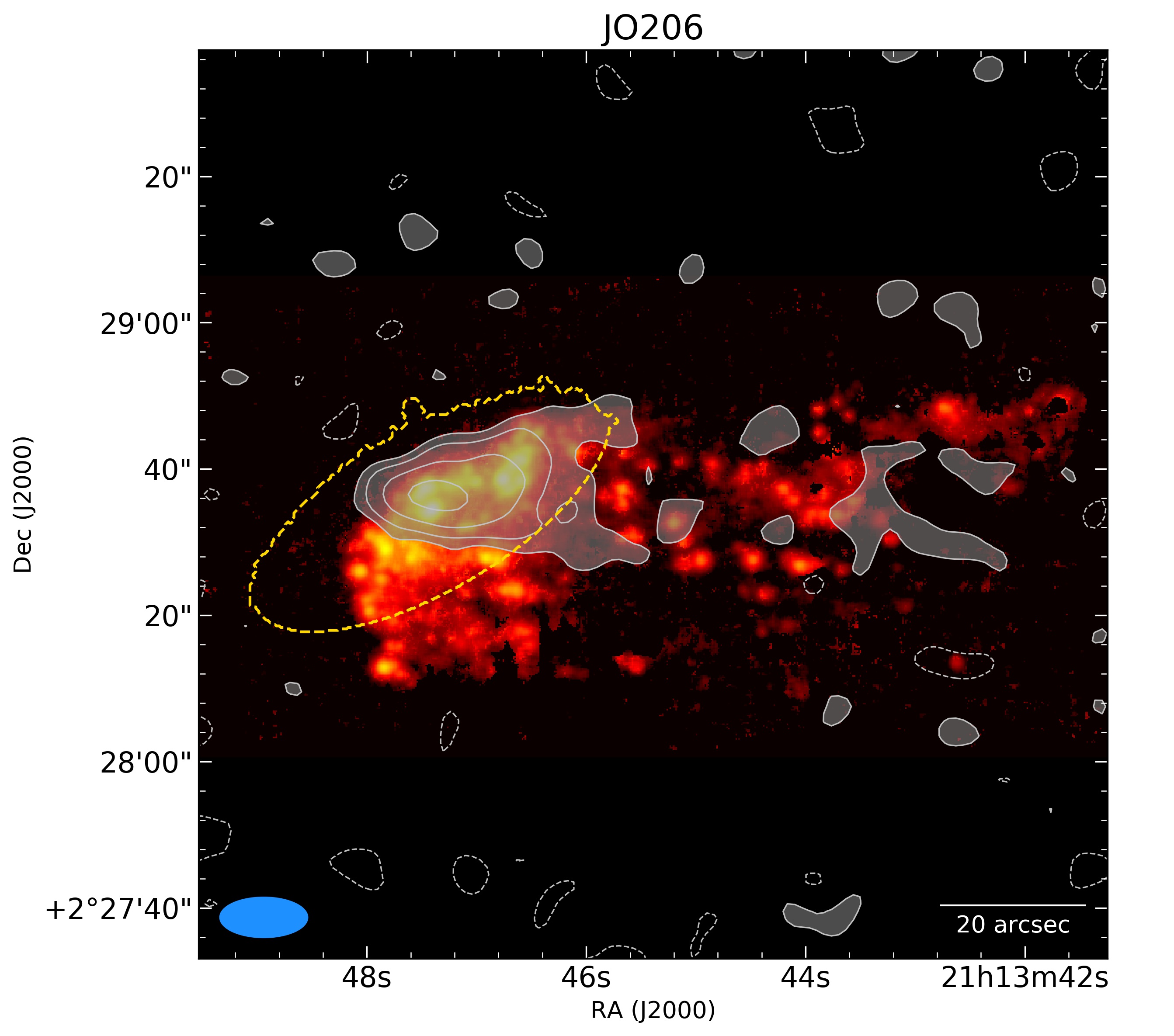}
\caption{\label{radio-halpha}MUSE H$\alpha$ images with the contours of the stellar continuum (gold dashed) and the LOFAR 144 MHz continuum emission (silver-filled contours). The radio continuum levels start from 2$\times$RMS level (except JW100 for which they start from 3$\times$RMS level) and progress by a factor 2.}
\end{figure*}

\begin{figure*}[t]
\includegraphics[width=.475\linewidth]{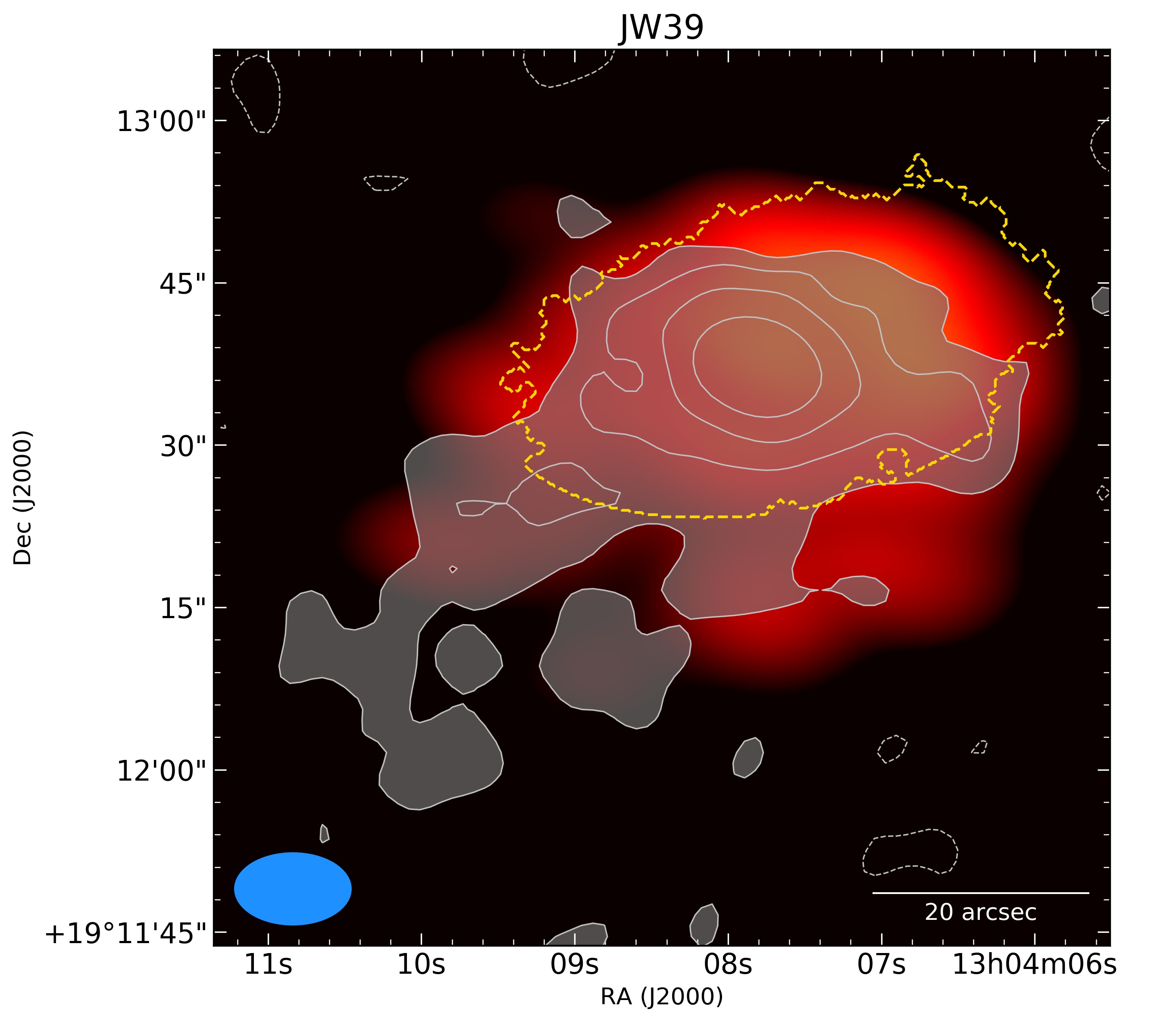}
\includegraphics[width=.475\linewidth]{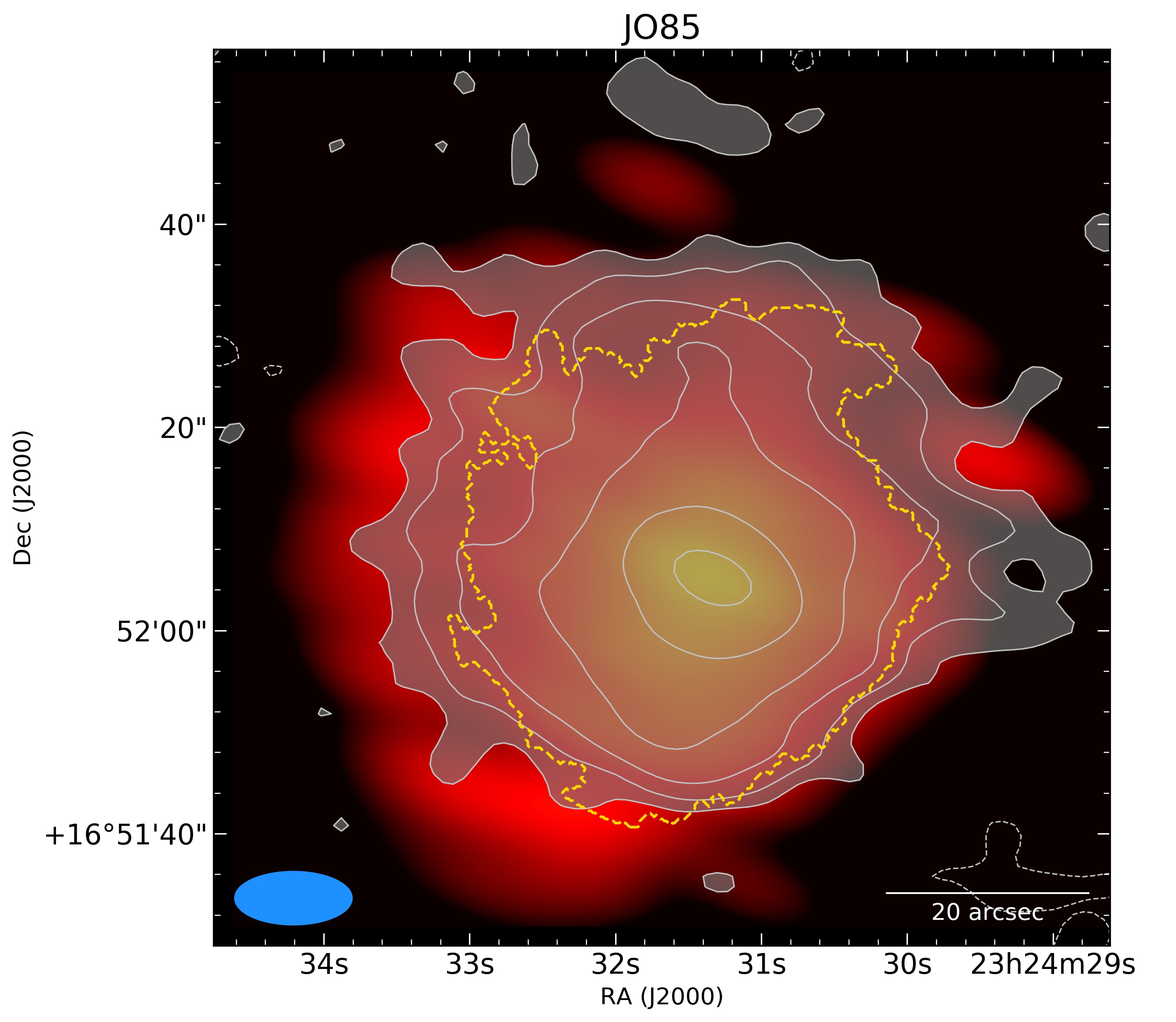}
\includegraphics[width=.475\linewidth]{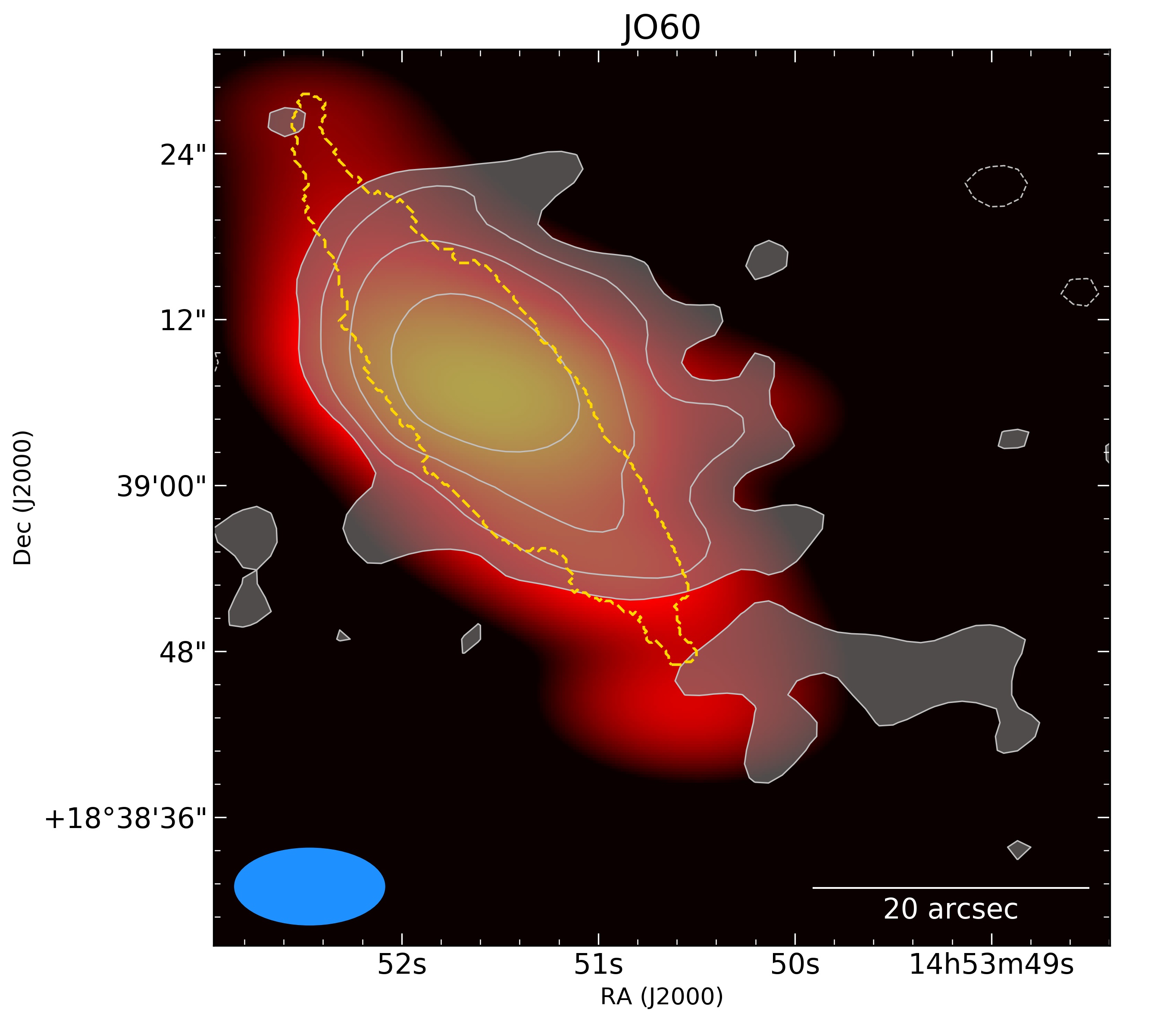}
\includegraphics[width=.475\linewidth]{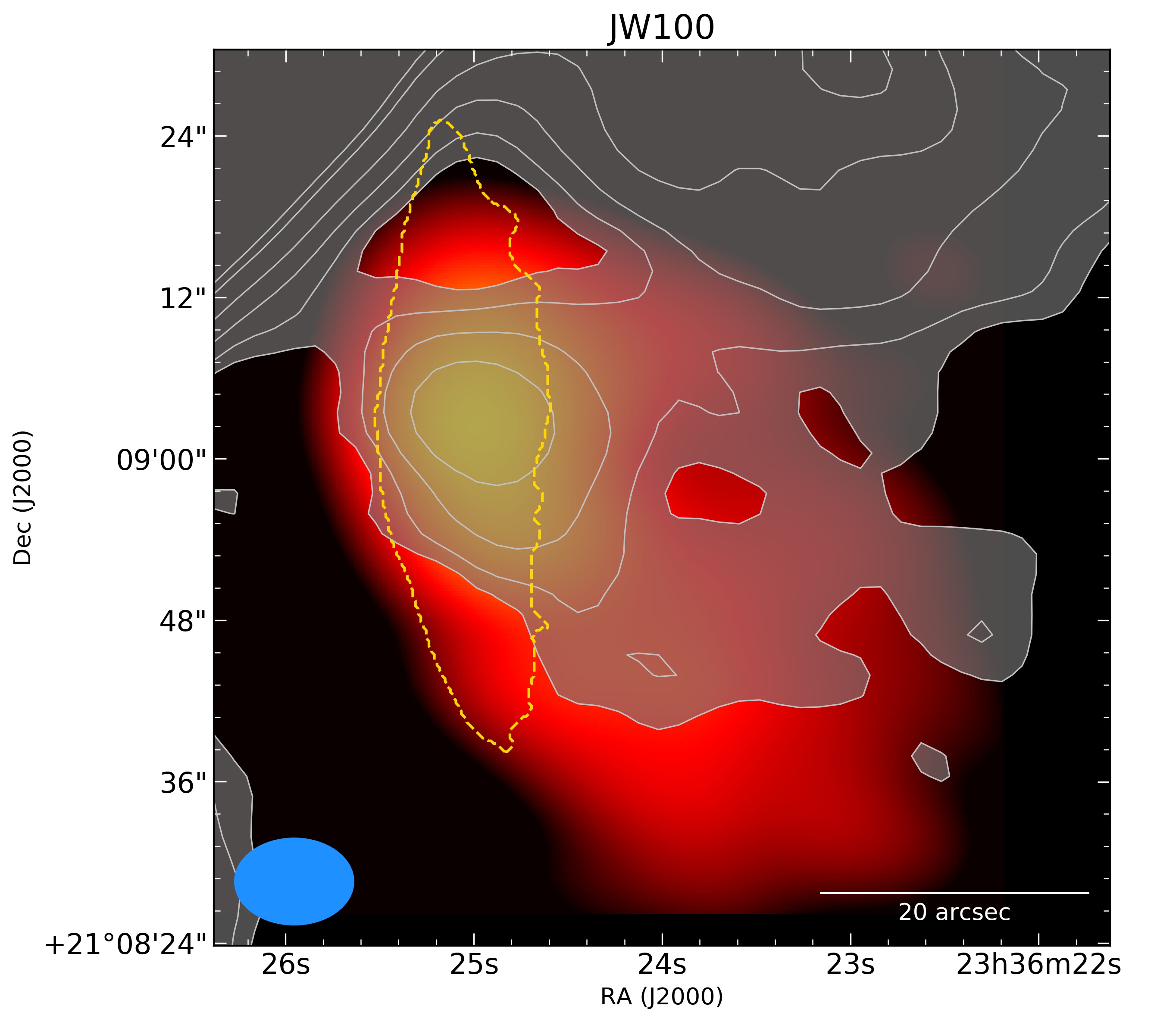}
\includegraphics[width=.475\linewidth]{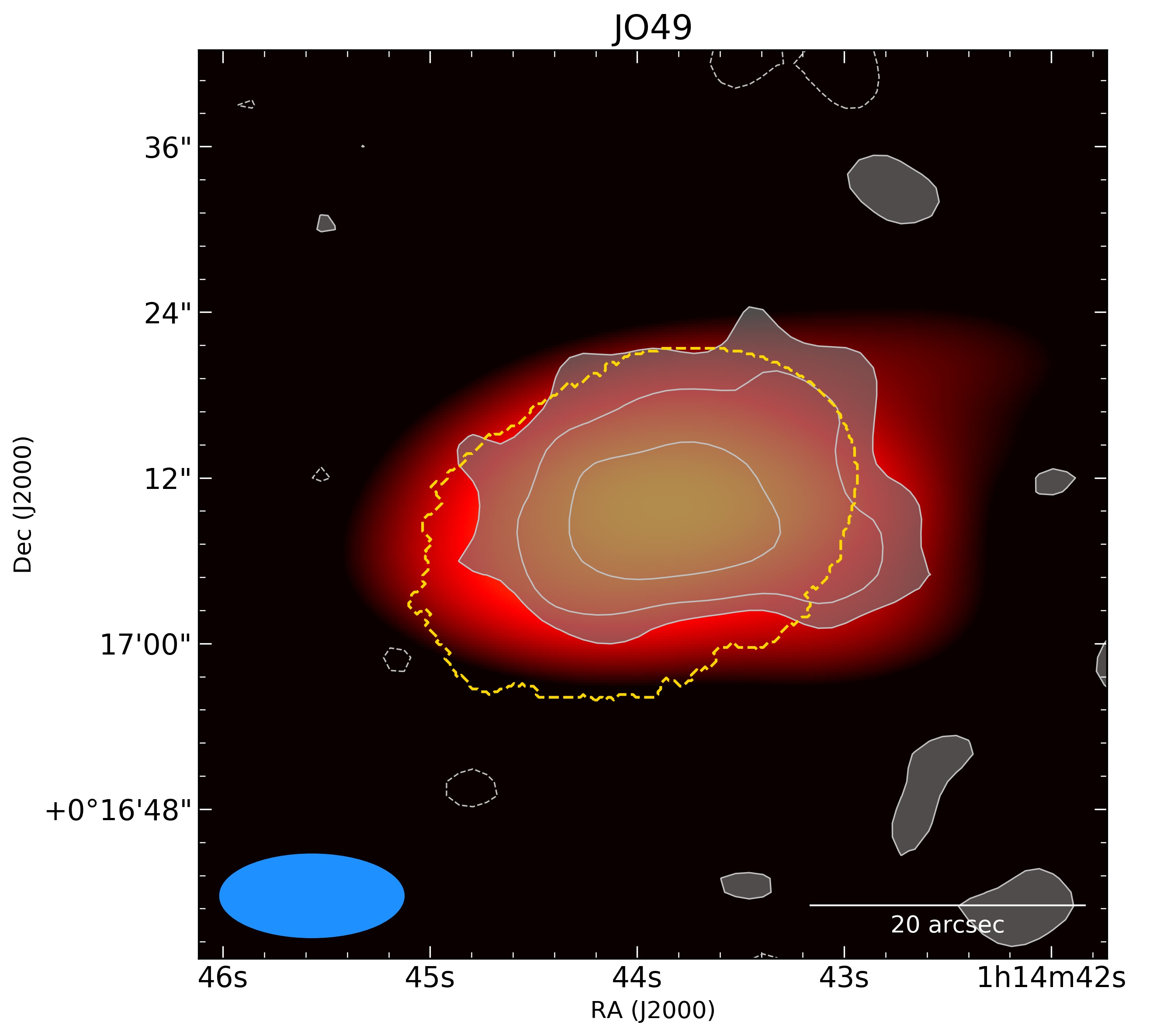}
\hfill\includegraphics[width=.475\linewidth]{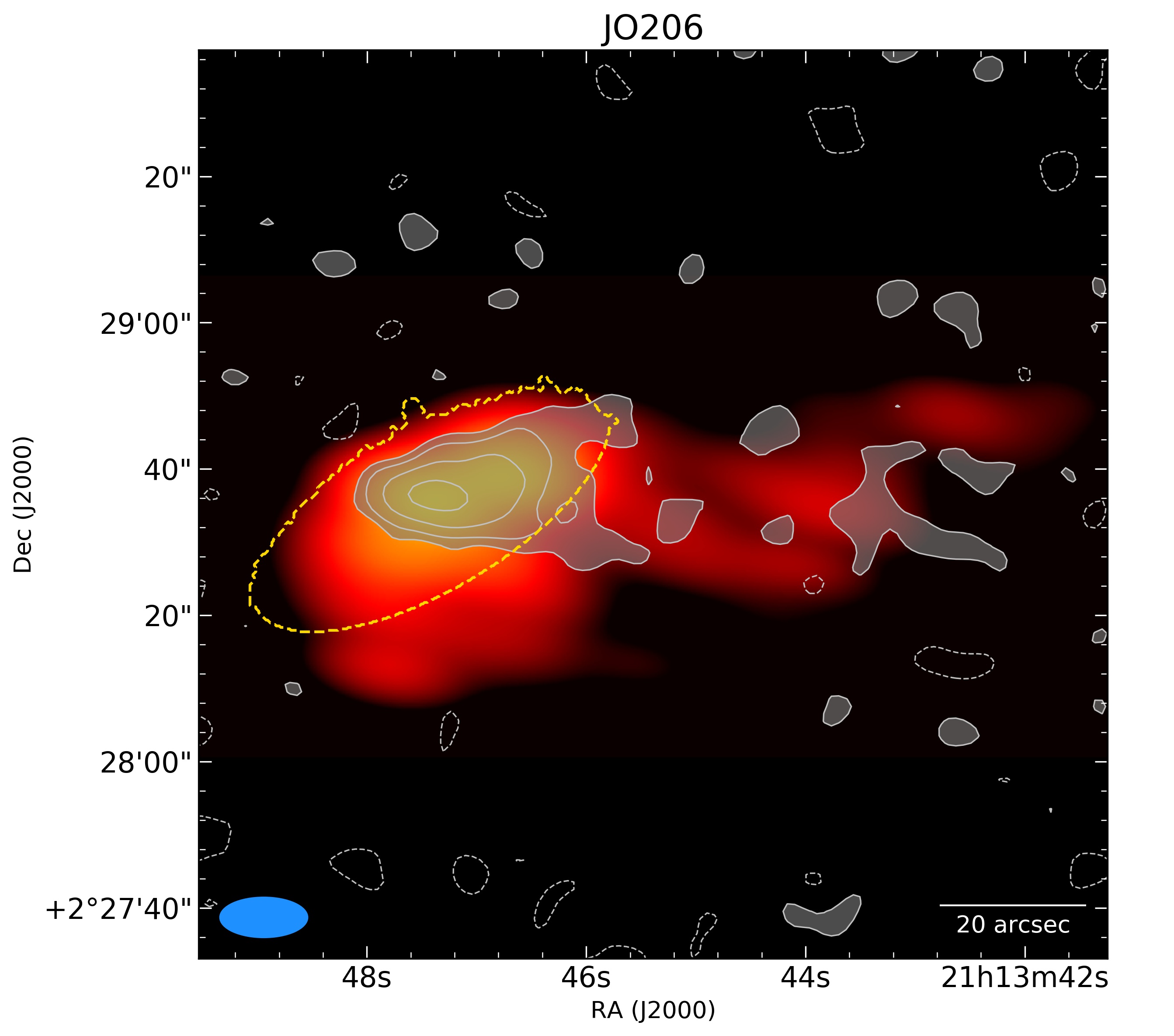}
\caption{\label{smooth}MUSE H$\alpha$ images smoothed to the resolution of the corresponding LOFAR images (see Table \ref{sample}) with the contours of the stellar continuum (gold dashed) and the LOFAR 144 MHz continuum emission (silver-filled contours). The radio continuum levels start from 2$\times$RMS level (except JW100 for which they start from 3$\times$RMS level) and progress by a factor 2.}
\end{figure*}

In order to investigate the connection between the stripped ISM and nonthermal radio emission, in Figure \ref{radio-halpha} we show the contours of the radio emission at 144 MHz of each galaxy (silver, filled) on top of their emission-only, dust-corrected H$\alpha$ emission. In Figure \ref{smooth} we present the same contours on top of the H$\alpha$ maps that have been smoothed  with a gaussian beam to match the resolution of the LOFAR images. In both figures we show also the contours of the stellar disk \citep[gold dashed,  from][]{Gullieuszik_2020}. The resolution and RMS noise level of each image is reported in Table \ref{sample}. We note that the emission to the north of JW100 (central-left panel in Figure \ref{radio-halpha}) is not related to the galaxy but it is instead part of the Kite radio source \citep[see][]{Ignesti_2020b,Ignesti_2021}. \\

The new LOFAR images reveal that in the stellar disk of these galaxies,  with the only exception of JO85, the radio emission is truncated with respect to the stellar emission (i.e., the radio emission is less extended than the stellar one, leaving radio-devoid regions at the outer edges of the stellar disk). In principle, this truncation may be the result of the  stripping of the ISM magnetic fields and the CRe, together with the ISM, or low-frequency flux losses occurring in the high-density, star forming regions of the disk due to the ionization captures. 
The former scenario should be characterized by lack of both radio and H$\alpha$ emission at the disk edges as result of the ISM stripping, whereas in the latter we should observe H$\alpha$ emission without radio counterpart and a local flattening of the radio spectral index at low frequencies due to the loss of the low-energy CRe \citep[e.g.,][]{Lacki_2013,Basu_2015,Heesen_2022}. In Figure \ref{radio-halpha} and \ref{smooth} we observe galaxies with signatures of stripping of both the thermal and nonthermal ISM (e.g., the northern edge of JW39), whereas evidence of ionization losses have been observed at the disk edges of JW100 in form of flat-spectrum emission between 144 MHz and 1.4 GHz  \citep[][]{Ignesti_2021}. Interestingly, in JO206 we observe both truncated radio and H$\alpha$ emission in the northern half of the disk, and H$\alpha$ without radio in the southern half. This suggests that the two processes, stripping and ionization captures, can coexist.\\

In general, the peak of the radio emission is close to the galactic centre. We evaluated the presence of radio-loud AGN by computing the radio-loudness parameter $\mathcal{R}:= (L_{\text{1.4 GHz}}/\text{W Hz}^{-1})/(L_{\text{H}\alpha}/L_\odot)$, where $L_{\text{1.4 GHz}}$, $L_{\text{H}\alpha}$ and $L_\odot$ are the radio at 1.4 GHz, H$\alpha$, and solar luminosities, respectively \citep[e.g.,][]{Kozie_2017}. To compute $\mathcal{R}$ we use $L_{\text{1.4 GHz}}$ measured in the NRAO VLA Sky Survey \citep[][]{Condon_1998}, and $L_{\text{H}\alpha}$ estimated from the MUSE observations by selecting the spaxels  classified as LINER or Seyfert in the BPT diagrams within a circular region centered on the galactic centre. A radio-loud AGN shows, typically, $\text{log}\mathcal{R}>15.8$, whereas in our sample we measure $13.1<\text{log}\mathcal{R}<14$, thus we conclude that the GASP-LOFAR sample does not contain radio-loud AGN. Nevertheless, we cannot rule out a contamination in the central emission due to low-power radio emission associated with AGN processes such as AGN driven wind, free-free emission from photo-ionized gas, low-power jets, or the black hole accretion  \citep[e.g.,][]{White_2017,Panessa_2019}.\\ 

The distribution of the radio emission with respect to the H$\alpha$ emission in the stellar disks is more diverse. Figure \ref{radio-halpha} shows galaxies in which radio and H$\alpha$ emission are similarly truncated with respect to the stellar disk (JW39 and JO49), or where radio emission appears to be more truncated than H$\alpha$ (JW100 and JO206), or where only radio is truncated whilst H$\alpha$ closely follows the stellar disk (JO60). JW39, JO49 and JO206 show radio-devoid regions located on their leading edge, that become more evident in the smoothed images (Figure \ref{smooth}). A similar feature has been observed also in the Virgo galaxies, and they have been interpreted as result of the ram pressure winds stripping the gaseous ISM  \citep[][]{Murphy2009}. At odds with the other galaxies, JO85 shows no radio-devoid regions within its stellar disk, although they are present on its eastern edge.  This sample shows a higher incidence of truncated disks at radio than H$\alpha$ emission which might  suggest that the nonthermal ISM components are more affected by the RPS than the warm plasma which, in turn, is more affected than the cold, star-forming gas because the ram pressure acceleration is inversely proportional to the gas column density \citep[e.g.,][]{Tonnesen_2021}. We suggest that the combination of this stripping `hierarchy' and ionization captures in the high-density gas can explain the different morphology of jellyfish galaxies at the different wavelength, and it will be further discussed in Section \ref{discu}.\\

JW39, JO60 and JW100 show unilateral, extraplanar radio emission extending for up to $\sim40$ arcsec in the direction of the stripped H$\alpha$ tail. JW39 shows also radio-devoid edges in the opposite direction. As observed in \citet[][]{Murphy2009}, the combination of these two features suggests that the relativistic, radio-emitting plasma could have been stripped from the disk into the tail. This provides very strong evidence that this emission is affected by the RPS. To a lesser extent, the radio emission appears to be aligned with the stripping direction in JO49. The extraplanar emission in JO206 appears to be composed of several, barely-resolved patches along the stripped H$\alpha$ tail, though, the detection is less certain. JO85 shows radio emission extending outside the stellar disk toward West, which may appear at odds with the fact that it is moving toward south-west \citep[][]{Bellhouse_2021}.

\subsection{The radio luminosity-star formation rate relation}
\label{sec1}
Cluster galaxies, and especially jellyfish galaxies, generally present an excess of radio emission with respect to their current SFR \citep[e.g.,][]{Gavazzi_1999,Roberts_2021a}. In order to test if the GASP-LOFAR galaxies follow this trend, we compare their radio luminosity at 144 MHz, including both the disk and the extraplanar emission, with their integrated SFR, measured in their disks and tails from the H$\alpha$ emission.  The integrated flux density of each galaxy, $S_\text{Rad}$, is evaluated within the lowest radio contour presented in Figure \ref{radio-halpha}. In order to evaluate the contribution of the extraplanar radio emission, we include the emission within the contours in Figure \ref{radio-halpha} 
that coincides with the H$\alpha$ tails. The corresponding error is the quadrature sum of the RMS contribution in the area used for the integrated intensity measure and the calibration error. We then compute the corresponding luminosity at 144 MHz, $L_{\text{Rad}}=4\pi D_L^2 S_\text{Rad}[(1+z)^{\alpha -1}]$, where $D_L$ is the luminosity distances of the hosting clusters (see Table \ref{sample})\footnote{ Due to the missing information on the spectral index of the galaxies, we prefer to neglect the $k$-correction. However, given the low redshift of the sample, its contribution is negligible.}. The resulting radio luminosity at 144 MHz are in the range $L_{\text{Rad}}=6-27\times10^{22}$ W Hz$^{-1}$. The integrated SFRs, including disks and tails, based on the H$\alpha$ emission (Table \ref{sample}) are  taken directly from \citet[][]{Vulcani2018}.\\

\begin{figure}
    \centering
    \includegraphics[width=\linewidth]{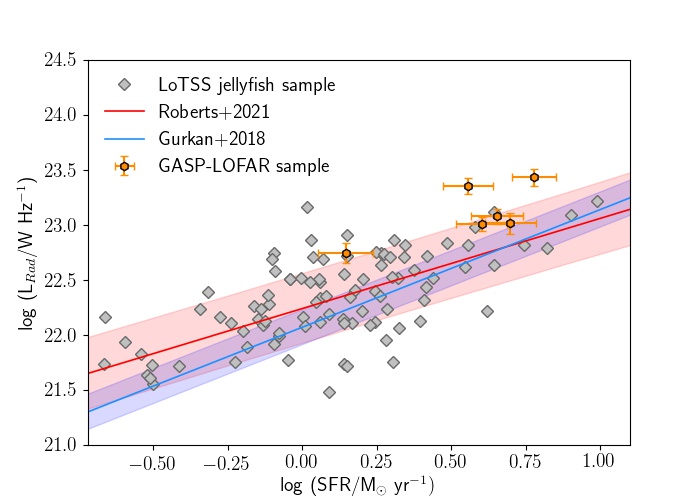}
    \caption{\label{lum-sfr}Radio luminosity at 144 MHz vs. SFR of the GASP-LOFAR (gold) and LoTSS (silver) samples. We report the empirical relations between $L_{\text{Rad}}$ and SFR presented in \citet[][]{Gurkan_2018} (blue) and \citet[][]{Roberts_2021a} (red), the latter is fitted on the LoTSS sample. The shaded regions show the dispersion at 1$\sigma$ around the corresponding best-fit relations.} 
\end{figure}

The values of $L_{\text{Rad}}$ and the integrated SFR are reported in Table \ref{sample} and Figure \ref{lum-sfr}. In the latter, we also show the values reported for 95 other LoTSS jellyfish galaxies from \citet[][]{Roberts_2021a}, and the corresponding best-fitting $L_\text{Rad}$-SFR relation at 144 MHz (red line). For reference, we show the empirical, global $L_{\text{Rad}}$-SFR relation computed at 144 MHz  presented in \citet[][]{Gurkan_2018} (blue line):  
\begin{equation}
\label{gurkan}
    \text{log }L_{\text{Rad}} =1.07 \pm0.08 \times\text{log SFR} +22.07\pm0.02\text{.}
\end{equation}
where $L_\text{Rad}$ and SFR are in units of W Hz$^{-1}$ and $M_\odot$ yr$^{-1}$, respectively. This relation was fitted on a sub-sample of $15,088$ star-forming galaxies with redshift $0<z<0.6$ from the seventh release of the Sloan Digital Sky Survey. The optical spectroscopic informations were provided by the $H$-ATLAS survey \citep[][]{Eales2010}.\\ 

The comparison with the previous studies highlights that the galaxies of the GASP-LOFAR sample generally occupy the poorly-sampled high SFR area. They show enhanced radio emission compared to the aforementioned relations (blue and red lines). They fall outside of the $1\sigma$ scatter of the \citet[][]{Gurkan_2018} relation (blue-shaded area) but, with the only exception of JO85 and JW39, within the larger scatter of the LoTSS Jellyfish's relation (red-shaded area). In the following, we further explore the origin of this radio excess by studying the connection between radio, H$\alpha$ emission and SFR.

\subsection{Point-to-point diagnostic diagrams}
\label{diag_desc}
The difference in resolution between the LOFAR and MUSE images does not permit a reliable comparison between the different maps on a pixel-by-pixel basis. Therefore, we used the point-to-point analysis\footnote{The point-to-point analysis was carried out with the Point-to-point TRend Extractor \citep[PT-REX, see][for further details]{Ignesti_2022}} to compare the two emissions over larger regions. For each galaxy, we sample the radio emission with a grid composed by rectangular cells large at least as the beam of each LOFAR image (Table \ref{sample}). Each cell covers a region with an average radio surface brightness above the levels discussed in Section \ref{lofi}. Then, we use those cells to simultaneously measure: 
\begin{enumerate}
\item Integrated radio flux density, $S_{\text{Rad}}$, with the corresponding errors. From this quantity we derive the radio surface brightness, $I_{\text{Rad}}$, (by dividing $S_{\text{Rad}}$ for the area of each cell) and luminosity, $L_\text{Rad}$, as done in Section \ref{sec1} for the total values;
\item Integrated, dust-corrected, emission-only H$\alpha$ flux density measured in  the smoothed H$\alpha$ maps (Figure \ref{smooth}). The uncertainties are defined as described in Section \ref{muse}. We compute the H$\alpha$ surface brightness, $I_{\text{H}\alpha}$, in each cell. Then, by using only the spaxels defined as star forming or composite in the BPT diagrams, we estimate the corresponding current SFR according to Equation \ref{kenny}, SFR$_{H\alpha}$;
\item Sum of the time-averaged SFR during the last  $2\times10^8$ (SFR$_{\text{recent}}$) yr prior to observations as reconstructed with the SINOPSIS code in each spaxel (see Section \ref{muse}). This time bin approximately represents the maximum radiative time of the CRe emitting at 144 MHz \citep[e.g.,][]{Basu_2015,Heesen_2019}. 

\end{enumerate}

\begin{figure*}
\includegraphics[width=0.5\linewidth]{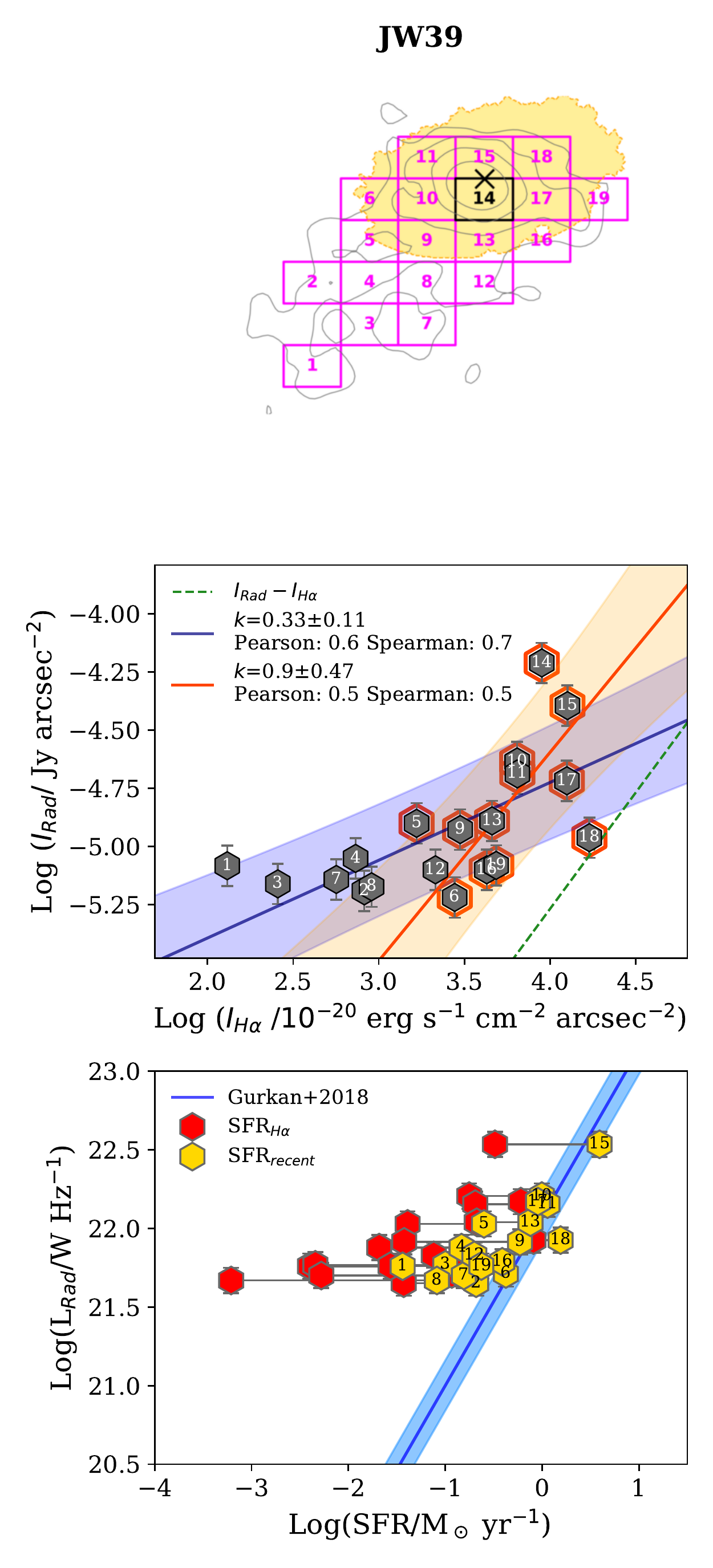}
\includegraphics[width=0.5\linewidth]{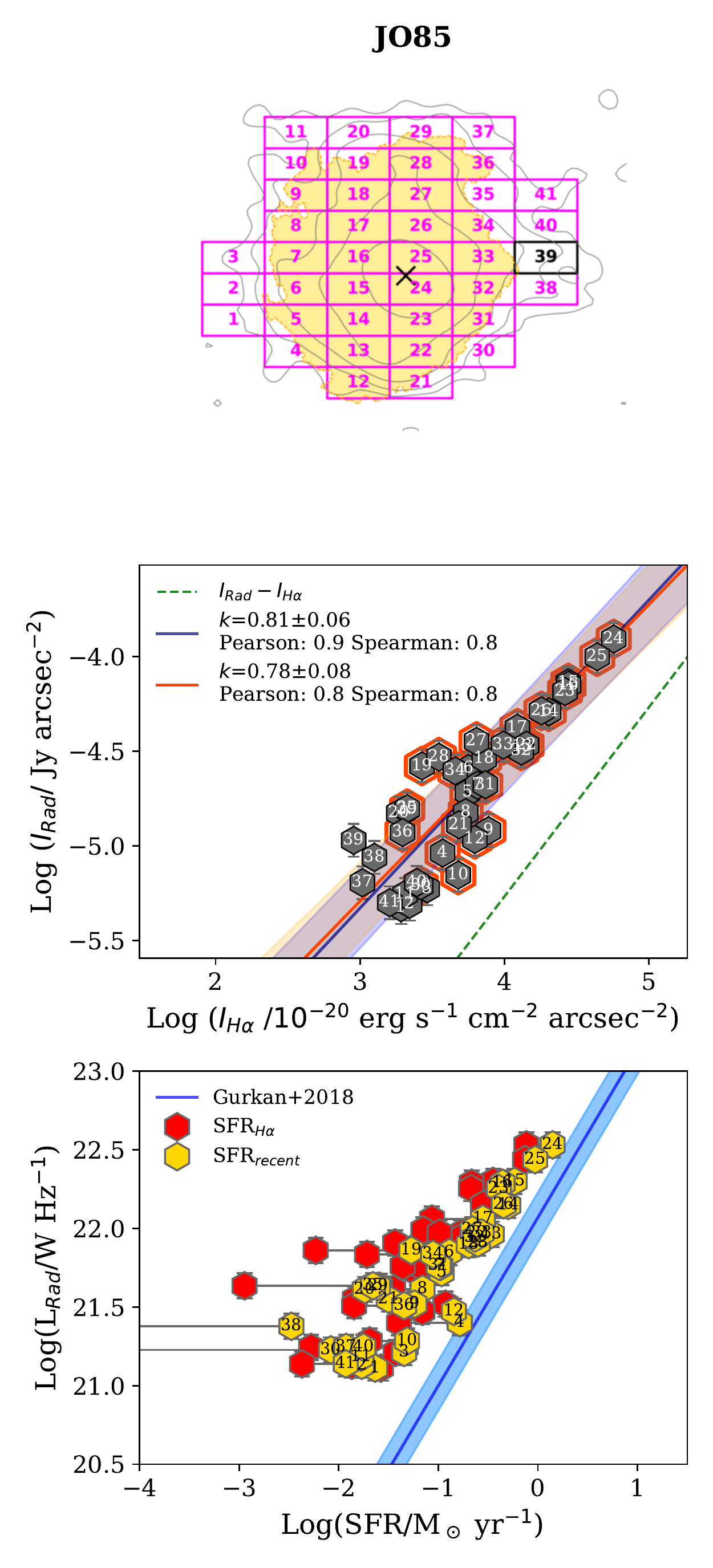}
\caption{\label{dp} Diagnostic plots for JW39 (left) and JO85 (right). Top: Sampling grid on top of the radio (grey) and the stellar disk (orange) contours. The star forming cells are reported in magenta. The numbers map the points in the other panels. The black cross shows the galactic centre; Centre: $I_{\text{Rad}}$ vs. $I_{\text{H$\alpha$}}$ measured from the maps in Figure \ref{smooth} with the best-fit (continuous, blue for the  total and orange for the disk-only) and the SFR-based (dashed, from Equation \ref{kenni-gurk}) relations. Bottom: $L_{\text{Rad}}$ vs. SFR (SFR$_{\text{recent}}$, gold, and SFR$_{H\alpha}$, red), with the trend expected from Equation \ref{gurkan} (blue line);  The shaded regions show the 1$\sigma$ confidence interval of the best-fit relations.} 
\end{figure*}

The aforementioned quantities are combined in a series of diagnostic plots shown in Figure \ref{dp}. For each galaxy, we show the sampling grid on top of the radio emission and stellar disk contours (top panel). We assign a number to each cell of the grid to ease locating the regions in the corresponding diagnostic plots. For reference, we indicate the position of the galactic centre with a black cross. The diagnostic plots include the radio vs. H$\alpha$ surface brightness (central panel, see Section \ref{p1}), and the radio luminosity vs. star formation rate (SFR$_{\text{recent}}$ and SFR$_{H\alpha}$) (bottom panel, see Section \ref{p2}).\\

In order to test if the grids provide a reliable picture of each galaxy, we tested the total radio vs. H$\alpha$ surface brightness correlation with the Monte Carlo point-to-point analysis (see Appendix \ref{MCptp}), finding that the general results did not depend on the choice of the grid. This indicates that the grids are fine enough to reliably sample the relevant spatial variations (i.e., $\sim10$ kpc at the galaxies' redshift), and, thus, that diagnostic plots are solid. However, we are aware that the point-point analysis is intrinsically biased by the arbitrary choice of the sampling grid, in the sense that the position of each point on the plots necessarily depends on the geometry of the grids. For this reason, in the followings we focus on the general results of the point-to-point analysis, instead of the individual features of each galaxy.

\renewcommand{\thefigure}{\arabic{figure} (continued)}
\addtocounter{figure}{-1}

\begin{figure*}[t]

\includegraphics[width=.5\linewidth]{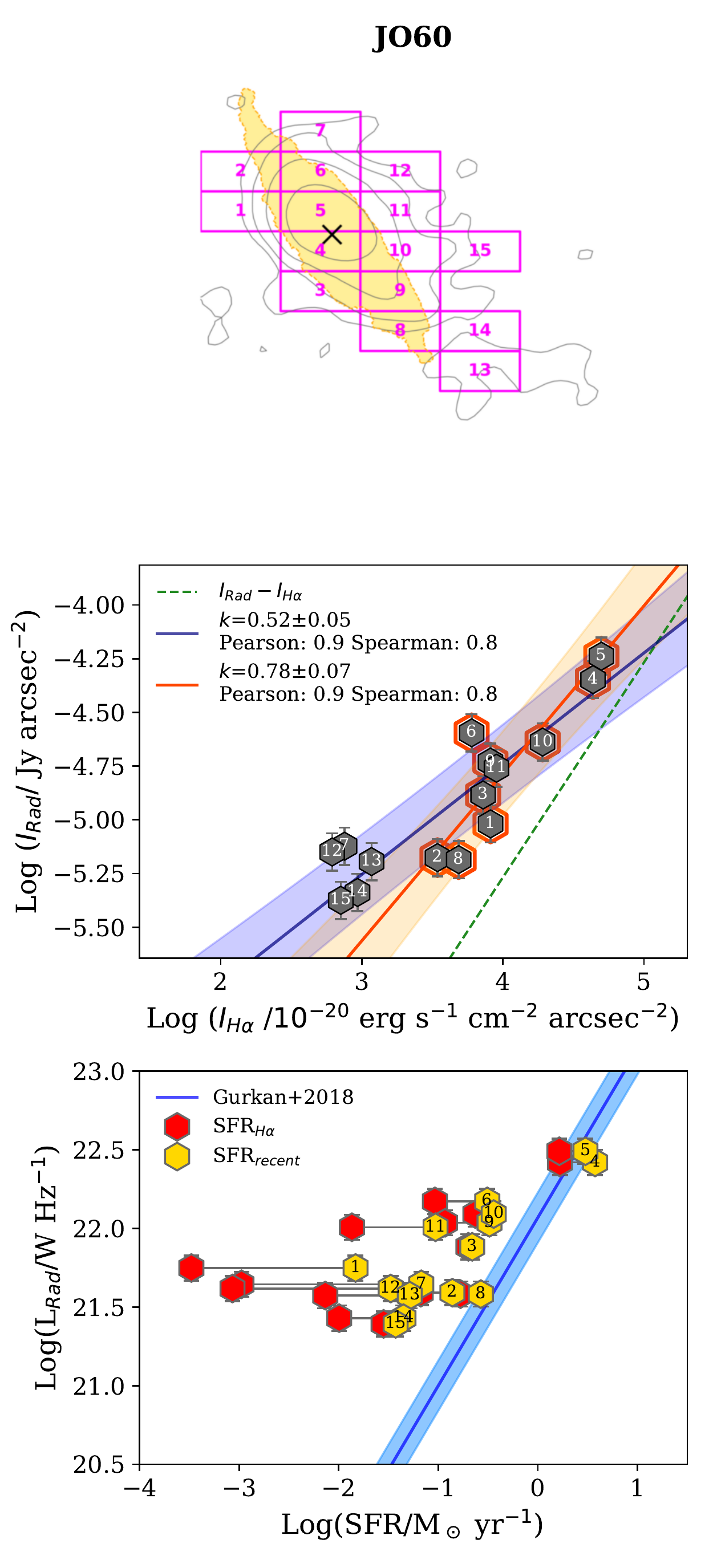}
\includegraphics[width=.5\linewidth]{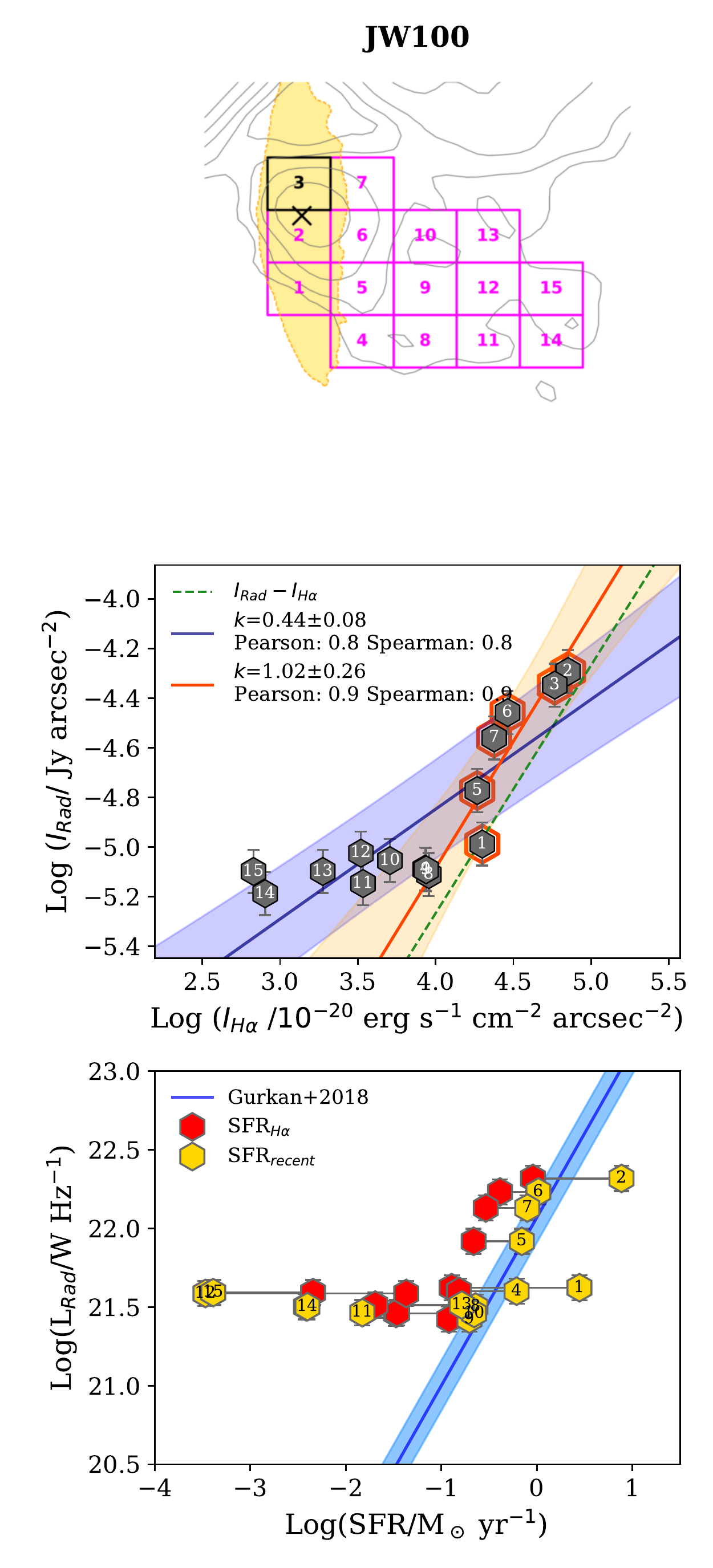}
\caption{Diagnostic plots for JO60 (left) and JW100 (right). }
\end{figure*}

\renewcommand{\thefigure}{\arabic{figure} (continued)}
\addtocounter{figure}{-1}

\begin{figure*}[t!]
\includegraphics[width=.5\linewidth]{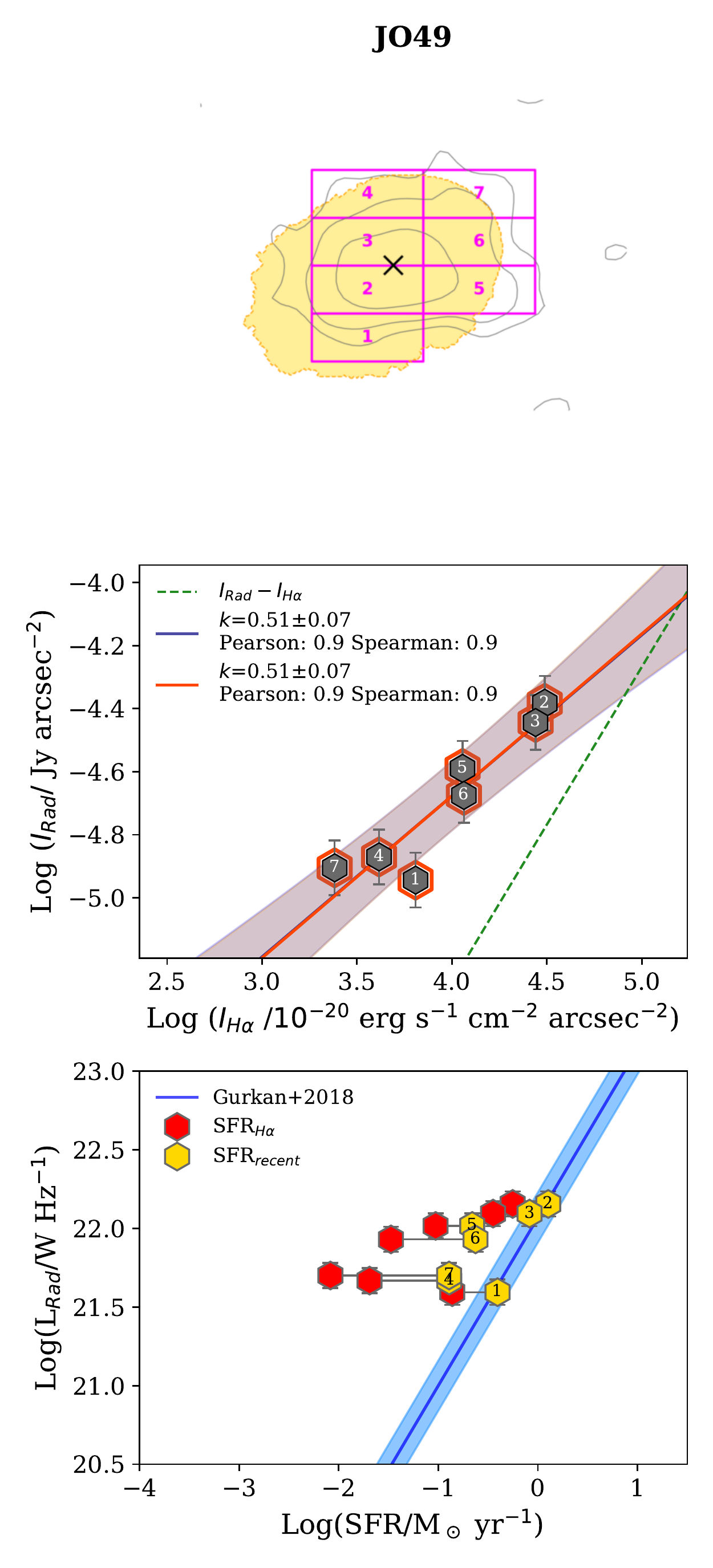}
\includegraphics[width=.5\linewidth]{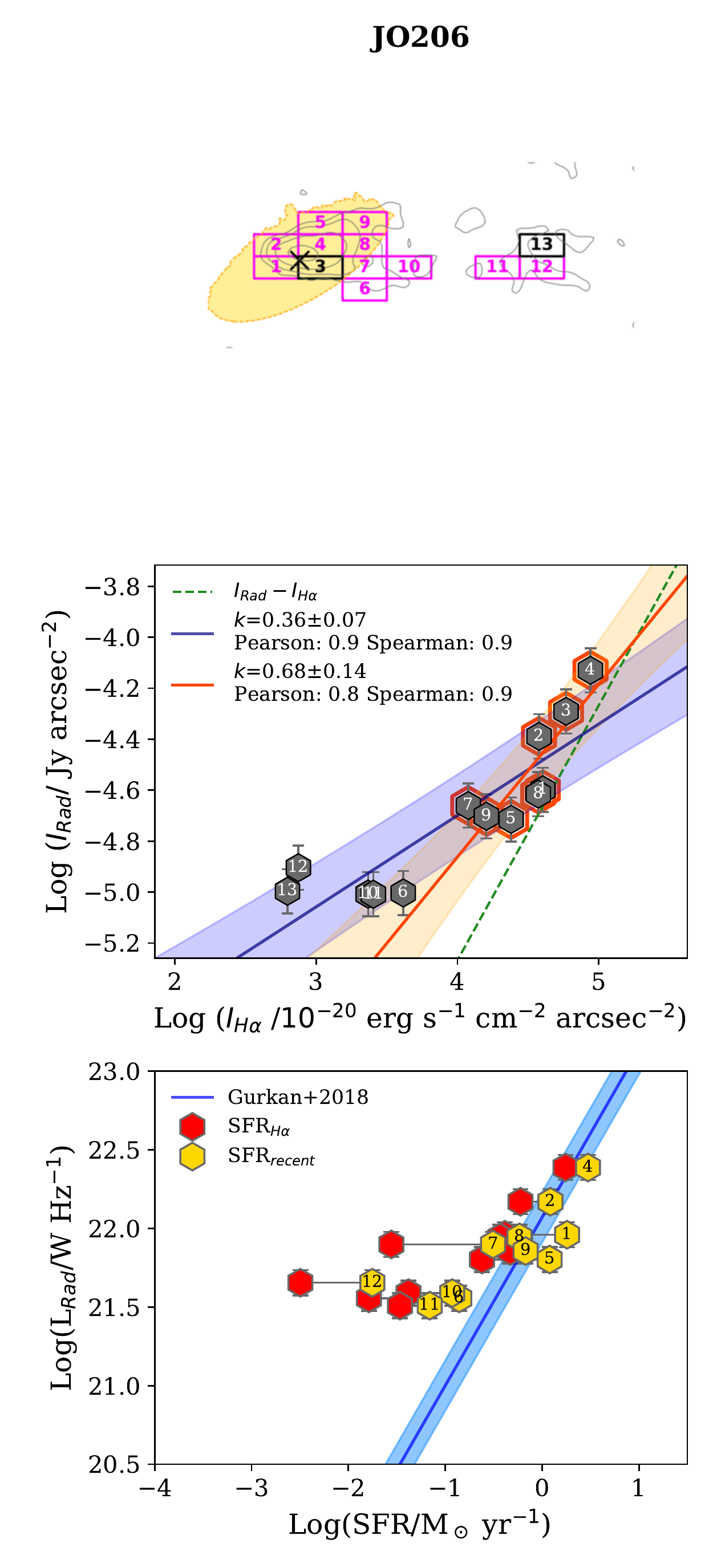}
\caption{Diagnostic plots for JO49 (left) and JO206 (right). }
\end{figure*}
\renewcommand{\thefigure}{\arabic{figure}}

\subsubsection{Radio vs. H$\alpha$ surface brightness} \label{p1}
\begin{figure}
    \centering
    \includegraphics[width=\linewidth]{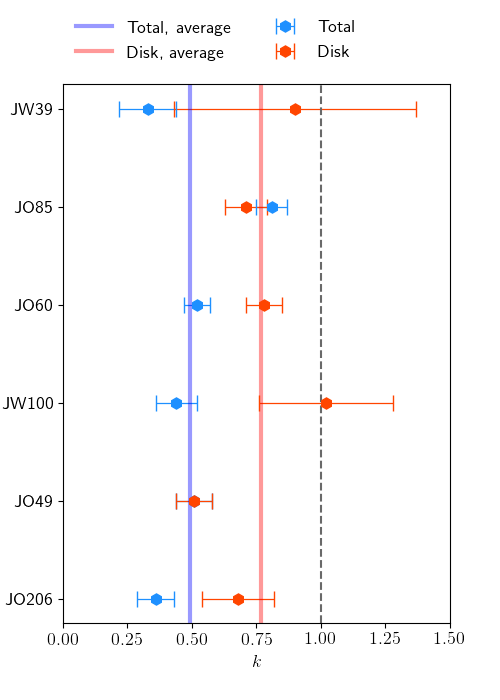}
    \caption{Slopes $k$ of the $I_{\text{Rad}}$-$I_{\text{H}\alpha}$ relations for each galaxy (Figure \ref{dp}). The vertical lines indicate the mean value of each distribution and the $k=1$ level (dashed grey line).}
    \label{k-dist}
\end{figure}

The comparison of the two surface brightness values can provide insights into the ISM microphysics and the connection between the different phases responsible for the two different types of  emission. In the central panels of Figure \ref{dp} we compare $I_\text{Rad}$ and $I_\text{H$\alpha$}$ computed for each cell.  We identify the disk cells as those that share $>10\%$ of their surface with the stellar disk mask shown in the top panel (gold contour), and indicate them with an orange contour. In this way we can infer the differences in the  $I_\text{Rad}-I_\text{H$\alpha$}$ spatial correlation between the total galaxy and the stellar disk-only. This corresponds to analyze how the spatial correlation between thermal and nonthermal ISM evolves from the stellar disk to the extraplanar regions. We report the errors on $I_\text{Rad}$ and $I_\text{H$\alpha$}$ derived from the respective fluxes (although the errors on $I_\text{H$\alpha$}$ are almost negligible and are smaller than the marker size). In order to test the presence of possible spatial correlations, both for the total galaxy and the  stellar disk only, we estimate the Spearman and Pearson correlation coefficients and fit a power-law $I_{\text{Rad}} \propto I_{\text{H$\alpha$}}^{k}$ (blue for the total galaxy and in orange for the disk) by using the bivariate correlated errors and intrinsic scatter method \citep[BCES,][]{Akritas_1996}. For reference, we plot also the global $I_\text{Rad}-I_\text{H$\alpha$}$ relation (green-dashed line) derived by combining Equations \ref{kenny} and \ref{gurkan}, and by converting the luminosity into surface brightness:
    \begin{equation}
    \frac{I_{\text{Rad}}}{\text{Jy arcsec}^2}\simeq5.4\times 10^{-10} \frac{I_\text{H$\alpha$}}{\text{erg s}^{-1}\text{cm}^{-2}\text{arcsec}^{-2}}.
    \label{kenni-gurk}
    \end{equation}\\
Equation \ref{kenni-gurk} describes the global  radio-to-H$\alpha$ ratio expected when both the emissions are powered by the total SFR. The resulting slopes $k$, for both the total and disk-only distributions are reported in the legend of each panel, and they are summarized in Figure \ref{k-dist}.\\

We find that, for every galaxy, the points lie above the global, linear correlation expected based on the standard SFR calibrations (Equation \ref{kenni-gurk}, dashed green line), which is expected from the global excess of radio emission of these galaxies (Figure \ref{lum-sfr}). This study reveals that a sub-linear spatial correlation holds between $I_{\text{Rad}}$ and $I_{\text{H$\alpha$}}$ in the form $I_{\text{Rad}} \propto I_{\text{H$\alpha$}}^{k}$ with an average $k=0.50$. In the stellar disks, the average slope steepens to $k=0.77$. The gap between the total and the disk-only slopes is more striking for those galaxies with the most evident radio tails, that are JW39, JO60, JW100, and JO206. Indeed in their corresponding plots we observe two distinct distributions, the disk cells which follow a steep, almost linear correlation, and those in the tail which show a flat trend with constant values of $I_{\text{Rad}}$ that flattens the total slopes. The implications of this result are discussed in Section \ref{spat}. As a caveat, we note that the selection of the disk cells, and hence the resulting slope, ultimately depends on our arbitrary choice of the overlap threshold ($>10\%$), that was chosen empirically. Due to the low number of cells used for the fit, increasing this threshold reduces the number of disk's cells and, thereby, the goodness of the fit. However, the main conclusion, that is the steepening of the slopes in the disk with respect to the total, does not change. 

\subsubsection{Radio luminosity vs. star formation rate} \label{p2}
In the bottom panels of Figure \ref{dp} we compare $L_{\text{Rad}}$ with SFR$_{\text{recent}}$ (gold points) and SFR$_{H\alpha}$ (red points), which trace the SFR in the last 2$\times10^8$ and $\sim10^7$ yr, respectively. This comparison permits us to test how much the radio excess depends on the fact that radio and H$\alpha$ emissions trace the SFR on different timescales \citep[][]{Kennicutt_2012}. For reference, we show also the global trend predicted by Equation \ref{gurkan} (blue line). As we are interested only in the star forming regions, we first select in the MUSE H$\alpha$ emission maps only those spaxels classified as star forming or composite in the BPT diagrams. Then  we arbitrarily define as 'star-forming', (magenta boxes in the top panels in Figure \ref{dp}) those cells in which $\geq50\%$ of the total H$\alpha$ flux computed in spaxels with a BPT classification comes from those classified as star-forming and composite, and we ultimately use only the star-forming cells for this analysis. \\

For every galaxy, the SFR$_{H\alpha}$ points are located above the expected $L_{\text{Rad}}-$SFR ratio, which is in agreement with their integrated radio luminosity excess (Figure \ref{lum-sfr}). However, the SFR$_{\text{recent}}$, which are generally higher than the SFR$_{H\alpha}$, are in better agreement with the expected ratio. This is observed both in the disks and the tails, which entails that this result does not depend on the threshold adopted to select the star forming cells. Increasing the threshold would only exclude those cells with less star-forming spaxels, which are usually located in the tails \citep[][]{Poggianti2019}. We also observe that the cells adjacent to the AGN show radio-to-SFR ratios that are in line with the rest of the galaxy, further suggesting that the contamination from the AGN radio emission is limited (see Section \ref{new_res}). This result is going to be further discussed in Section \ref{eccesso_sfr}.


\section{Discussion}
\label{discu}

\subsection{Implications of the $I_{\text{Rad}}$-$I_{\text{H$\alpha$}}$ spatial correlation}
\label{spat}
The jellyfish galaxies in the GASP-LOFAR sample show sub-linear $I_\text{Rad}$-$I_{\text{H$\alpha$}}$ spatial correlations, which are at odds with the linear, global one expected based on the standard SFR-calibrators (Equation \ref{kenni-gurk}). Moreover, all the points are located above this relation, which indicates that the radio-to-H$\alpha$ excess is ubiquitous in these galaxies. These results suggest that the nonthermal radio emission does not depend solely on the local, ongoing star formation. The galaxies with the most extended extraplanar emission (JW39, JO60, JW100 and JO206) show different trends in the total's and disk's slope, with the slopes measured in their disks being systematically steeper than those derived by using the whole galaxy (Figure \ref{k-dist}).  Moreover, the more distant, faintest regions have, on average, a constant $I_{\text{Rad}}$ and higher ratios of radio-to-H$\alpha$ emission with respect to Equation \ref{kenni-gurk} than those in the disk. \\
 
The existence of sub-linear spatial correlations between the resolved nonthermal radio emission and the ISM star-formation tracers in spiral galaxies has been previously reported in literature  \citep[e.g.,][]{Hoernes_1998,Berkhuijsen_2013,Heesen_2019}. In general,  the correlation slope can depend on the observed radio frequency, where the lower the frequency, the flatter the correlation, or on the radio spectral index, where the steeper the spectral index, the flatter the slope \citep[][]{Heesen_2019}. Those pieces of evidence suggest that CRe diffusion plays a role in defining these spatial correlations and, more specifically, that the flattening of the slope is a consequence of the CRe transport away from the star forming regions. This would entail that old CRe, which can be traced by low-frequency observations, should be the main responsible for the spatial correlations flattening because they have been transported over large scales during their radiative time. In this framework, the sub-linear spatial correlations observed in the disks of GASP-LOFAR sample (Figure \ref{k-dist}) could be explained by a combination of radio emission dominated by the old, steep-spectrum CRe, which is consistent with the steep-spectrum nonthermal radio emission observed in JO206 and JW100 \citep[][]{Muller_2021, Ignesti_2021}, and an efficient CRe transport.\\

For JW39 and JW100 the disks' slopes are consistent with the linearity, that is the trend expected from Equation \ref{kenni-gurk}. This may suggests that the smoothing scale ($\sim10$ kpc) is similar to the CRe transport scale, and, hence, the slopes resembles the standard SFR-calibrators. Therefore, the spatial correlation slopes observed in the disks of JW39 and JW100 are consistent with the idea that the CRe transport scale in these galaxies is larger than observed in spiral galaxies' disks ($\sim 1-5$ kpc). We investigate this result by comparing the spatial scales, $L$, of the two principal CRe transport processes in these systems, that are diffusion and advection. In the case of CRe diffusion, the spatial scale is:
\begin{equation}
    L=\sqrt{4D\tau}\simeq 1.1\times \sqrt{\frac{D}{10^{28}\text{cm}^2\text{s}^{-1}}\frac{\tau}{\text{10 Myr}}}\text{ kpc,} 
    \label{diff}
\end{equation}
where $\tau$ is a time scale, and $D$ is the diffusion coefficient, which depends on the CRe energy and the local magnetic field power spectrum, and can vary between $10^{27}$ and $10^{29}$ cm$^2$ s$^{-1}$ \citep[][]{Strong_2007}. For the CRe advection, the typical spatial scales is: 
\begin{equation}
L=V\cdot\tau\simeq  \frac{V}{\text{100 km s}^{-1}}\frac{\tau}{\text{10 Myr}}\text{ kpc,}
\label{adv}
\end{equation}
where $V$ is the CRe velocity.  Equation \ref{diff} shows that to reach $L=10$ kpc, for a typical $D=1\times10^{28}$ cm$^2$ s$^{-1}$, the time scale is of the order of $\sim$900 Myr, which is longer than the typical CRe radiative time in galactic disks. Covering these spatial scales in $\leq10^8$ yr requires a diffusion coefficient $D\geq9\times10^{28}$ cm$^2$ s$^{-1}$, which is slightly higher than observed in spiral galaxies \citep[e.g.,][]{Strong_2007, Tabatabaei_2013, Heesen_2019}. On the other hand, these scales would be consistent with the CRe advection (Equation \ref{adv}), as a typical velocity of 100 km s$^{-1}$ would be able to cover 10 kpc in $\sim10$ Myr, which is more reasonable for low-energy electrons emitting at 144 MHz. Therefore, our results might hint that CRe transport in the disks of these jellyfish galaxies, JW39 and JW100, is either dominated by advection, due to the ram pressure which is stripping the nonthermal ISM (see Section \ref{new_res}), or that the CRe diffusion might be more efficient than in normal galaxies. The diffusion coefficient $D$ depends on the local magnetic field configuration and turbulence spectrum \citep[][]{Strong_2007}, thus it may be possible that the RPS, by affecting the ISM microphysics, may induce higher values of $D$ and, hence, a more efficient CRe diffusion (Equation \ref{diff}) than the one observed in normal spiral galaxies. However, we note that, for the rest of the sample, the spatial correlations in their stellar disks are not consistent with the linearity (Figure \ref{k-dist}). We argue that this might be due to projection effects that mix the disk and the extraplanar emissions, which, on the basis of what we observe for JW39, JO60, JW100, and JO206, follows a flat, almost uniform, distribution. Thus this blend  may result in a flattening of the disks' slopes. Another possible explanation could be a discrepancy between the sampling resolution and the transport scale that was not solved by the smoothing. \\

Concerning the extraplanar emission, observing a sub-linear $I_{\text{Rad}}- I_{\text{H$\alpha$}}$ relation might shed some light on the physics of the stripped ISM. Previous studies  reported that jellyfish galaxies commonly show a steepening of the radio spectral index in the extraplanar emission \citep[][]{Vollmer_2004,Muller_2021,Ignesti_2021,Roberts_2021c} that indicates that the CRe in these regions were originally accelerated in the disk and subsequently swept in the tail by the ram pressure wind. In the case of the radio tails observed in GASP-LOFAR sample, which show projected lenghts of $\sim40$ kpc, Equation \ref{adv} indicates that a ram pressure wind speed $V\geq400$ km s$^{-1}$ would be required to transport the CRe in $\leq10^{8}$ yr. For comparison, for magnetic field intensities $B\leq10$ $\mu$G \citep[][]{Ignesti_2021}, and CRe energy losses dominated by synchrotron (i.e., $U_{\text{Rad}}/U_{\text{B}}\simeq 0$), Equation \ref{t-syn} yields $t_{\text{syn}}\geq9\times10^7$ yr for CRe emitting at 144 MHz in the tails. Concerning the wind speed, albeit the total velocity of these galaxies is still unknown, the line-of-sight velocities for JW39, JO60, JW100, and JO206 are 820, 1000, 1950, and 632 km s$^{-1}$, respectively \citep[][]{Gullieuszik_2020}. Therefore, the observed radio tails lengths are consistent with a pure advection scenario. Our results provide complementary pieces of evidence for this scenario. Observing that the extraplanar regions locate the farthest from the trend defined by Equation \ref{kenni-gurk} can be naturally explained by the presence of additional electrons coming from the disk that flatten the spatial correlation slope by boosting the local radio emission in the extraplanar regions with the faintest H$\alpha$ emission. 
\subsection{The echoes of the past star formation}
\label{eccesso_sfr}
\begin{figure}
    \centering
    \includegraphics[width=\linewidth]{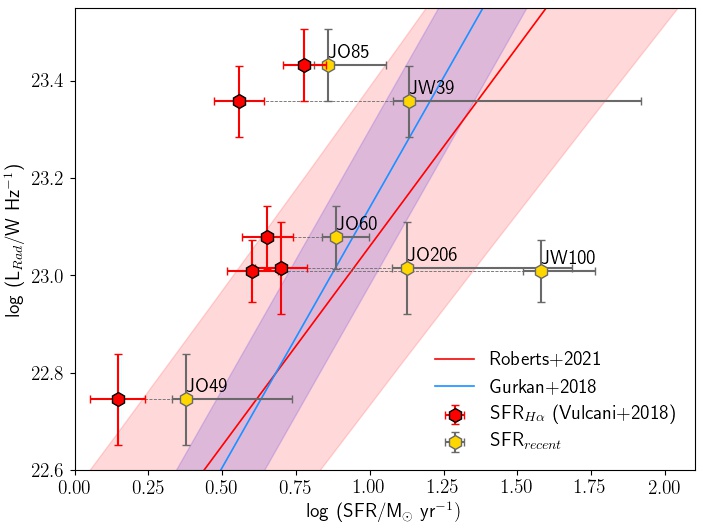}
    \caption{Integrated radio luminosity at 144 MHz vs. SFR. For each galaxy, the  SFR$_{H\alpha}$ from \citet[][]{Vulcani2018} (red points) and SFR$_{\text{recent}}$ (gold points, see Section \ref{diag_desc}), with the corresponding errors, are connected by the dashed lines. We show also the best-fit $L_{\text{Rad}}-$SFR relations from \citet[][]{Gurkan_2018} (blue line) and \citet[][]{Roberts_2021a} (red line). The shaded regions show the dispersion at 1$\sigma$ around the corresponding best-fit relations. }
    \label{l-sfr}
\end{figure}
The GASP-LOFAR galaxies, analogously to cluster galaxies, show an excess of radio emission with respect to the ongoing SFR (Figure \ref{lum-sfr}). As highlighted by the $L_{\text{Rad}}$-SFR plots (Section \ref{p2}), this excess does not appear to be localized on the leading edge of the galaxies \citep[as observed in][]{Murphy2009}, but it is 
ubiquitous over the galaxies. Historically, this radio excess has been explained as the results of ISM compression, or due to the contribution of the CRe re-accelerated by the ICM winds \citep[e.g.,][]{Gavazzi1995,Boselli2006, Murphy2009}. In \citet[][]{Ignesti_2021} we presented an alternative view based on the multi-frequency analysis of JW100. By comparing the excess of SFR between the radio and H$\alpha$ estimates with the average SFR inferred by the stellar population synthesis, we found the radio excess in its stellar disk to be consistent with a decline of the SFR happening within the time-scale of the radio emission (i.e., $10^{7-8}$ yr). This scenario was further supported by the fact that the nonthermal emission in both the disk and the tail of JW100 showed evidence of spectral steepening above 1.4 GHz, thus indicating that the radio emission is dominated by old CRe. \\

In this work we further explore the hypothesis of a connection between the radio luminosity excess and the SFR decline. In the $L_\text{Rad}$ vs. SFR point-to-point plots (Section \ref{p2}) we observe that the SFR$_{\text{recent}}$, which is generally higher than the SFR$_{H\alpha}$, is close to the expected ratio (blue line in Figure \ref{dp}, bottom plots), especially for the central regions. In the extraplanar regions the excess seems to persist, which would be consistent with the idea of a local radio emission boost provided by the CRe advected from the disk by the ram pressure winds. In Figure \ref{l-sfr} we present the integrated $L_{\text{Rad}}$ and SFR$_{H\alpha}$ for the GASP sample (red points, see Table \ref{sample}) and the corresponding integrated SFR$_{\text{recent}}$ (gold points). For every galaxy  SFR$_{\text{recent}}>$SFR$_{H\alpha}$, which indicates that the SFR in these galaxies declined in the last $2\times10^8$ yr. The extent of the SFR decline varies within the sample, ranging from a factor $\sim10$ in JW100 to $\sim2$ in JO85.\\

Overall, we observe that SFR$_{\text{recent}}$ is in better agreement than SFR$_{H\alpha}$ with the global, empirical relations \citep[blue and red lines, from][]{Gurkan_2018,Roberts_2021a}. Specifically, we measure a decrease of the root-mean-square deviations from 0.5 to 0.4 and from 0.4 to 0.3 for the \citet[][]{Gurkan_2018} and \citet[][]{Roberts_2021a} relations, respectively, as more galaxies fit within the $1\sigma$ scatter (Figure \ref{l-sfr}). At odds with the other galaxies, the total radio emission of JW100 appears to be more in agreement with the current SFR rather than the past one, whereas in the stellar disk the radio emission is well explained by the past star formation \citep[][]{Ignesti_2021}.\\

The result shown in Figure \ref{l-sfr} supports the idea that the current radio emission is related to the past, higher SFR, (or, in other words, that the radio emission is dominated by old CRe) and thus, that the jellyfish galaxies' radio excess is a consequence of the fast evolution of the star formation activity induced by the RPS in these galaxies. Qualitatively, a past phase of higher SFR would entail a higher CRe injection, thus a higher radio emission. When the SFR declines, the CRe injection should decline correspondingly. But if the decline time-scale is fast ($\sim10^7$ yr) , such as in the case of those galaxies heavily processed by RPS, and  shorter than CRe radiative time, its effect on the total radio emission (especially at lower frequencies) goes unseen because the bulk of the radio emission is produced by the old CRe injected previously. On the contrary, tracers which probe the most recent SFR (e.g., H$\alpha$) are affected by the SFR decline \citep[e.g.,][for a study of the SFR evolution with multiple tracers]{Cleland_2021}. Thus the result is that the radio emission appears in excess with respect to the current SFR.\\

Observations at higher frequency will be able to conclusively test this scenario. Observing signatures of a old, steep-spectrum CRe population (i.e., spectral steepening at frequencies above 1 GHz) would confirm that the radio emission is dominated by the older CRe injected during a past phase of higher SFR.  Indeed, evidence of steep-spectrum emission at high frequencies have already been reported for JW100, which shows an integrated, nonthermal spectral index  $\alpha=-0.92\pm0.15$ and $-2.43\pm0.37$ in the $1.4-3.2$ and $3.2-5.5$ GHz bands, respectively, in the disk, and $\alpha<-0.85$ between 1.4 and 3.2 GHz in the tail \citep[][]{Ignesti_2021}, and for JO206, which shows $\alpha\simeq-1$ in the disk and $\alpha\simeq-2$ in the tail between 1.4 and 2.7 GHz \citep[][]{Muller_2021}. Moreover, by comparing the detection levels of JO206's radio tail in the new LOFAR data and in the maps presented in \citet[][]{Muller_2021}, we can now roughly constrain the spectral index between 144 MHz and 2.7 GHz to $\alpha=-1.1$, which further indicates that the synchrotron spectrum is curved, i.e. that the CRe population is old. In order to conclusively test this scenario, the study of spectral index in the GASP-LOFAR sample will be the object of future works. This scenario should also imply that the radio excess broadly depends on the observed radio frequency. Observations at higher frequencies trace CRe with higher energy and shorter radiative times, hence more similar to the other tracers and more sensible to the SFR decline. On the contrary, observations at lower frequencies, such as the ones that are going to be provided by the LOFAR LBA Sky Survey \citep[][]{deGasperin_2021}, should observe an higher radio excess because they trace CRe with the longer radiative times. This scenario can be further explored also by detailed numerical simulations of these galaxies to trace the CRe evolution by taking into account the different transport process and the different energy loss mechanism in a multi-phase ISM.\\

Our conclusions do not rule out that the other processes proposed previously in literature (such as magnetic field amplification due to ISM compression, or CRe re-acceleration, see Section \ref{sec:intro}) could further contribute in producing the jellyfish galaxies' radio luminosity excess. However, we note that the results in Figure \ref{dp} challenge the scenario of radio excess due to ISM compression \citep[][]{Boselli2006}. In the $I_\text{Rad}$ vs. $I_{\text{H}\alpha}$ plots we observe that the radio excess is not limited to the regions in the compressed zone (i.e., where the contours of the radio emission are more dense) and, on the contrary, in some cases we observe that these regions, which should be the most compressed by RPS, are the closest to the expected radio-H$\alpha$ ratio. This could result from multiple effects in action. It could be that in those regions, due to the increased density of the compressed ISM, the low-energy electrons emitting at 144 MHz are more affected by the ionization captures \citep[e.g.,][]{Basu_2015}, thus balancing the synchrotron emissivity enhancement due to RPS. It could be that those regions, as a consequence of the ISM compression, are harbouring an intense phase of ongoing star formation that dominates the local radio emission and, therefore, they are less affected by the echoes of the past star formation. Furthermore, we note that, due to the radio images resolution, the sampling cells located on the leading edges of some of these galaxies include both radio-rich and -deficient star-forming regions (Figure \ref{smooth}). This blend may result into a seemingly regular radio-to-SFR ratio.  

\section{Summary and conclusions}
We presented a study of the extended, nonthermal radio emission at 144 MHz for a sample of six jellyfish galaxies from the GASP sample. The GASP-LOFAR sample shows high radio luminosities ($6-27\times10^{22}$ W Hz$^{-1}$), and a series of recurring patterns in the radio and H$\alpha$ morphology, such as truncated radio emission with respect to the stellar disk and extraplanar components. These results highlight that the nonthermal ISM components, that are CRe and magnetic fields, are heavily influenced by the RPS similarly to the warm, ionized ISM. We observe an excess of radio luminosity in the GASP-LOFAR sample with respect to their current SFR. This is in agreement with the previous studies of jellyfish galaxies, and cluster galaxies in general. We devise a series of diagnostic plots to frame the connection between ISM stripping, star formation history, and radio emission. We found that a sub-linear relations holds between the radio and H$\alpha$ surface brightness ($I_{\text{Rad}}\propto I_{\text{H}\alpha}^k$) with an average slope  $k=0.50$, whereas in the stellar disks we measure $k=0.77$. The extraplanar regions have, in general, higher relative deviance from the expected radio luminosity-SFR relation than those in the disk. We argue that this is due to the presence of CRe coming from the disk that boost the local radio emission.\\

We observe that for JW39 and JW100 the $I_{\text{Rad}}-I_{\text{H}\alpha}$ correlation in their stellar disks is consistent with being linear, which would agree with the trend expected from the SFR calibrators. We speculate that it is because the smoothing scale is consistent with the CRe transport scale. In this case, a spatial scale of $\sim10$ kpc would entail that the CRe transport in their stellar disks is more efficient than observed in normal spiral galaxies. We speculate that this large spatial scale could be explained either by a CRe transport dominated by advection rather than diffusion, or a diffusion efficiency higher than usual induced by RPS.\\

From the comparison between the radio luminosity and the recent SFR evolution, both local and integrated, it emerges that  the SFR of these galaxies declined in the last $10^8$ yr, and that the radio luminosity can be explained better by the past, higher SFR than by the ongoing one traced by the H$\alpha$ emission. The implications are twofold. On the one hand, our work shows that the combined radio-H$\alpha$ studies can constrain the SFR history. On the other hand, the radio luminosity excess could emerge when the timescale of the SFR decline is shorter than the CRe radiative time, leaving a population of old CRe that produces the bulk of the radio emission. The CRe advection by ram pressure outside of the disk would then be the responsible of the extraplanar radio emission. This connection between star formation history and radio luminosity might hint that, in general, the integrated radio luminosity excess of cluster galaxies could be signature of a (fast) SFR decline. This work is among the few studies comparing the resolved properties of H$\alpha$ and low frequency radio emission, and is surely focused on a peculiar class of objects. However, the intimate connection between the recent, instead of the ongoing, star formation and the radio emission could be a general property of cluster galaxies that have a similar star formation histories. Future studies on larger samples of galaxies, including those not evidently affected by the ram-pressure, will help clarifying the possible caveats in using the low-frequency radio emission as a tracer for the ongoing SFR.\\

The results presented here call now for follow-up observations at higher frequencies. By combining these new pieces of information with detailed spectral index maps it will be possible to study the CRe lifecycle within these galaxies. Moreover, our findings provide a benchmark for numerical simulations of RPS, or more in general of cloud-crushing scenarios. The fact that, thanks to LOFAR, we are discovering more and more jellyfish galaxies with extraplanar, nonthermal radio emission indicates that large-scale magnetic fields are common in these systems, and that their evolution is intimately connected with the evolution of the stripped gas.   
\section*{Acknowledgments}
We thank the Referee for the constructive report that improved the presentation of this work. This work is the fruit of the collaboration between GASP and the LOFAR Survey Key Project team (``MoU: Exploring the low-frequency side of jellyfish galaxies with LOFAR", PI A. Ignesti). We acknowledge the INAF founding program 'Ricerca Fondamentale 2022' (PI A. Ignesti). A.I., B.V., R.P., M.G. acknowledge the Italian PRIN-Miur 2017 (PI A. Cimatti). This project has received funding from the European Research Council (ERC) under the European Union's Horizon 2020 research and innovation programme (grant agreement No. 833824). R.J.vW and A.B. acknowledges support from the VIDI research programme with project number 639.042.729, which is financed by the Netherlands Organisation for Scientific Research (NWO). I.D.R.  acknowledges support from the ERC Starting Grant Cluster Web 804208. S.L.M. acknowledges support from the Science and Technology Facilities Council through grant number ST/N021702/1. J.F. acknowledges financial support from the UNAM- DGAPA-PAPIIT IN111620 grant, México. We acknowledge funding from the INAF main-stream funding programme (PI B. Vulcani). \\

Based on observations collected at the European Organization for Astronomical Research in the Southern Hemisphere under ESO programme 196.B-0578.  LOFAR \citep[][]{vanHaarlem_2013}  is the Low Frequency Array designed and constructed by
ASTRON. It has observing, data processing, and data storage facilities in several countries,
which are owned by various parties (each with their own funding sources), and that are
collectively operated by the ILT foundation under a joint scientific policy. The ILT resources
have benefited from the following recent major funding sources: CNRS-INSU, Observatoire de
Paris and Université d'Orléans, France; BMBF, MIWF-NRW, MPG, Germany; Science
Foundation Ireland (SFI), Department of Business, Enterprise and Innovation (DBEI), Ireland;
NWO, The Netherlands; The Science and Technology Facilities Council, UK; Ministry of
Science and Higher Education, Poland; The Istituto Nazionale di Astrofisica (INAF), Italy.
This research made use of the Dutch national e-infrastructure with support of the SURF
Cooperative (e-infra 180169) and the LOFAR e-infra group. The Jülich LOFAR Long Term
Archive and the German LOFAR network are both coordinated and operated by the Jülich
Supercomputing Centre (JSC), and computing resources on the supercomputer JUWELS at JSC
were provided by the Gauss Centre for Supercomputing e.V. (grant CHTB00) through the John
von Neumann Institute for Computing (NIC).
This research made use of the University of Hertfordshire high-performance computing facility
and the LOFAR-UK computing facility located at the University of Hertfordshire and supported
by STFC [ST/P000096/1], and of the Italian LOFAR IT computing infrastructure supported and
operated by INAF, and by the Physics Department of Turin university (under an agreement with
Consorzio Interuniversitario per la Fisica Spaziale) at the C3S Supercomputing Centre, Italy. This research made use of Astropy, a community-developed core Python package for Astronomy \citep[][]{astropy_2013, astropy_2018}, and APLpy, an open-source plotting package for Python \citep[][]{Robitaille_2012}. A.I. thanks the music of Black Sabbath for providing inspiration during the preparation of the draft.
\bibliography{bib}{}
\bibliographystyle{aasjournal}

\appendix

\section{Monte Carlo point-to-point analysis}
\label{MCptp}
\begin{figure}[h]
    \centering
    \includegraphics[width=.6\linewidth]{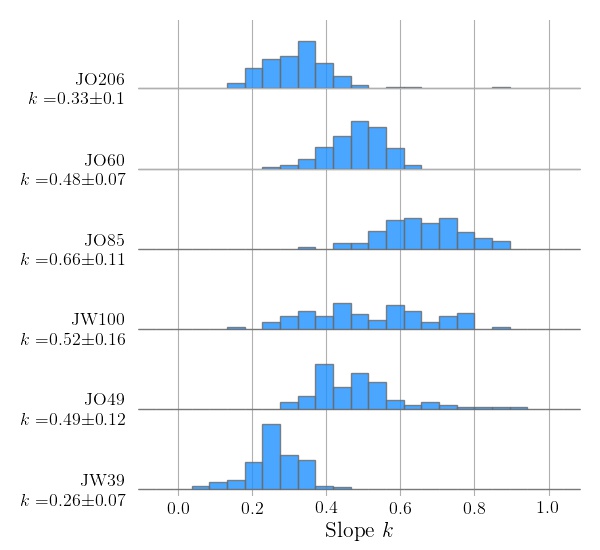}
    \caption{\label{mcp}Results of MCptp with 100 iterations for each galaxy of the GASP-LOFAR sample. For each galaxy, the mean and the standard deviation of the distribution is reported in the respective label.}
    \label{mcptp}
\end{figure}

We tested the reliability of the grids adopted in the point-to-point analysis (Figure \ref{dp}) by performing a Monte Carlo point-to-point analysis \citep[MCptp, for further details see][]{Ignesti_2022}. For each galaxy, we replied 100 different sampling of the radio emission with randomly-generated grids (each grids with the same cell size and thresholds of the ones shown in Figure \ref{dp}) and measured the slope $k$ of the corresponding  $I_{\text{Rad}}-I_{\text{H$\alpha$}}$ correlation. Then, by bootstrapping a value of $k$ from the best-fit of each cycle, we composed a distribution of 100 slopes for each galaxy. In Figure \ref{mcp} we report the final distributions and the mean and standard deviation of the slope $k$ for each galaxy. The values reported in Figure \ref{dp} fall within these distributions. This supports that the results presented in Section 3 are not biased by the choice of the grids.     


\end{document}